\documentclass[fleqn,usenatbib]{mnras}

\usepackage{newtxtext,newtxmath}

\usepackage[T1]{fontenc}
\usepackage{ae,aecompl}
\usepackage{multirow}
\usepackage{url}

\usepackage{graphicx}	
\usepackage{amsmath}	
\usepackage{float}

\DeclareMathOperator*{\argminA}{arg\,min}

\title[Inferring the properties of the reionization sources]{Inferring the properties of the sources of reionization using the morphological spectra of the ionized regions}

\author[S. Gazagnes et al.]{
Simon Gazagnes,$^{1,2,3}$\thanks{E-mail: s.r.n.gazagnes@rug.nl}
L\'eon V.E. Koopmans,$^{2}$
and Michael H.F. Wilkinson$^{3}$
\\
$^{1}$Energy and Sustainability Research Institute Groningen, University of Groningen, PO Box 11103, 9700 CC Groningen, The Netherlands\\
$^{2}$Kapteyn Astronomical Institute, University of Groningen, P.O Box 800, 9700 AV Groningen, The Netherlands\\
$^{3}$Bernoulli Institute for Mathematics, Computer Science and Artificial Intelligence, University of Groningen, P.O Box 407, \\9700 AK Groningen, The Netherlands 
}

\date{Accepted XXX. Received YYY; in original form ZZZ}

\pubyear{2020}

\begin{document}
\label{firstpage}
\pagerange{\pageref{firstpage}--\pageref{lastpage}}
\maketitle

\begin{abstract}
High-redshift 21-cm observations will provide crucial insights into the physical processes of the Epoch of Reionization. Next-generation interferometers such as the Square Kilometer Array will have enough sensitivity to directly image the 21-cm fluctuations and trace the evolution of the ionizing fronts. In this work, we develop an inferential approach to recover the sources and IGM properties of the process of reionization using the number and, in particular, the morphological pattern spectra of the ionized regions extracted from realistic mock observations. To do so, we extend the Markov Chain Monte Carlo analysis tool  \textsc{21CMMC} by including these 21-cm tomographic statistics and compare this method to only using the power spectrum. We demonstrate that the evolution of the number-count and morphology of the ionized regions as a function of redshift provides independent information to disentangle multiple reionization scenarios because it probes the average ionizing budget per baryon. Although less precise, we find that constraints inferred using 21-cm tomographic statistics are more robust to the presence of contaminants such as foreground residuals. This work highlights that combining power spectrum and tomographic analyses more accurately recovers the astrophysics of reionization.
\end{abstract}




\section{Introduction}
\label{sect:intro}

The formation of the first stars and galaxies through gravitational instability of small density fluctuations, several hundred million years after recombination, marks the beginning of a key phase transition in the Universe history referred to as the Cosmic Dawn. These first light sources emitted photons with enough energy to ionize the neutral hydrogen, which propagated in the Intergalactic medium (IGM) and progressively reionized the entire Universe. This latter era is called the Epoch of Reionization (EoR), where the transition from the Cosmic Dawn is rather ill-defined \citep[see, e.g][for reviews]{ciardi2005, morales2010, pritchard2012, furlanetto2016_reio, dayal2018}. Many unknowns remain about the nature of theses sources, and  especially their efficiency in re-ionizing the surrounding neutral-hydrogen environment. Over the past decade, both observational and theoretical studies have suggested that a population of low-mass, very star-forming galaxies with an average escape fraction of ionizing photons of $f_{\rm esc} = 0.1-0.2$, were the dominant suppliers of ionizing photons during the EoR \citep{ouchi, bouwens2012, robertson, dressler, finkelstein2019,mason2019, trebitsch2020}. Nevertheless, the contribution of brighter sources is still actively discussed \citep[][]{fontanot2012, fontanot2014, robertson2015, madau2015, mitra2018, naidu2020, dayal2020}. Further observational studies are required to constrain the escape fraction of these sources during reionization. Directly measuring $f_{\rm esc}$ at high redshifts is very challenging because of the limited mean free path of these photons. Hence, indirect proxies are needed to estimate the ionizing efficiency of the sources of reionization, and some progress has been recently made in that direction using analogs at lower redshift \citep[e.g][]{schaerer2016, verhamme2017, chisholm2018, chisholm2020, henry2018, wang2019, cen2020, gazagnes2018, gazagnes2020, izotov2020}.

The detection of the 21-cm line resulting from the spin-flip of the neutral Hydrogen electron can provide complementary  information to investigate the nature of ionizing sources and the propagation of ionizing fronts. First-generation interferometers, such as the Murchison Widefield Array\footnote{http://www.mwatelescope.org} \citep[MWA;][]{tingay2013, bowman2013}, the Low-Frequency Array\footnote{http://www.lofar.org} \citep[LOFAR;][]{vanhaarlem2013}, the Giant Metrewave Radio Telescope (GMRT)\footnote{http://www.ncra.tifr.res.in/ncra/gmrt} \citep{swarup1991}, or the Precision Array for Probing the Epoch of Reionization\footnote{http://eor.berkeley.edu} \citep[PAPER;][]{parsons2010}, aim at detecting the 21-cm fluctuations during the EoR. While no detection has been achieved yet, they yield tremendous improvements in our understanding of interferometric observations. Recently, improved upper limits  on a statistical detection of the 21-cm fluctuations (at the 2$\sigma$ level) using the power spectrum have been published in the literature: \citet{trott2020}  reported $\Delta^2_{21} \leq$  (43 mK)$^2$ at k = 0.14 h Mpc$^{-1}$ and $z$ = 6.5 using 110 hours of MWA observations  and  \citet{mertens2020} found  $\Delta^2_{21} \leq$ (73 mK)$^2$ at k = 0.075 h Mpc$^{-1}$ and $z$ = 9.1 using 141 hours of LOFAR observations. The latter results already put some light on the astrophysics of reionization by ruling out exotic scenarios incompatible with this upper limit \citep{greig2020_lofar, ghara2020_lofar, mondal2020}.

The progress with current instruments, but also the development of the second-generation interferometers such as the Hydrogen Epoch of Reionization Array\footnote{https://reionization.org/} \citep[HERA;][]{deboer2017} and the Square Kilometre Array\footnote{https://www.skatelescope.org/} \citep[SKA;][]{mellema2013, koopmans2015} will progressively tighten our understanding of the EoR by either yielding deeper limits on the power spectrum of the 21-cm fluctuations, or directly imaging it with redshift. Additional insights can be extracted from the non-Gaussian part of the 21-cm fluctuations, which are not encoded in the power spectrum. Over the past decades, a large number of theoretical studies have introduced new statistical formalism to extract information from the 21-cm signal.   \citet{shimabukuro2016}, \citet{majumdar2018}, \citet{watkinson2019} or \citet{hutter2020} have used the 21-cm bispectrum and shown that it could provide valuable insights about the ionization topology and the size distribution of the ionized structures. Similarly, \citet{gorce2019} have shown that comparable insights could be extracted using 3-points correlations functions. \citet{watkinson2014} and \citet{banet2020} investigated the skewness and kurtosis of the 21-cm intensity probability distribution function (PDF) and have shown that these non-Gaussianity measurements can trace the astrophysics of reionization. 

Finally, HERA and SKA should have enough sensitivity to map the 21-cm fluctuations and provide images of the evolution of the ionizing fronts during the EoR. The use of these 21-cm tomographic images has already been shown useful to analyze the size statistics of the ionized or neutral regions \citep{kakiichi2017, giri2018_bsd, giri2019neutral}, the morphology of the brightness temperature fluctuations \citep{chen2019, kapahtia2019}, or the topology of the ionizing field \citep{elbers2019, giri2020betti}. All these studies have  emphasized that alternative approaches to the power spectrum could provide very valuable insights to understand the astrophysics of cosmic reionization, and could be  used to break the degeneracy between models with similar power spectra \citep{kakiichi2017}. 

However, further analysis is required to assess how these approaches can be incorporated within a 21-cm analysis tool to quantify their robustness in inferring the astrophysical information lying in the 21-cm fluctuations. Recently, several studies explored the use of deep-learning approaches to infer the astrophysics of the process of reionization using 21-cm images \citep[e.g][]{gillet2019, hassan2019}. In this paper, we investigate how independent statistics extracted from these images can be included in a Bayesian statistical inference framework. To do so, we extend 21CMMC \citep{greig2015, greig2017}, a  Markov Chain Monte Marco (MCMC) analysis tool, originally designed to quantify reionization properties using the power spectrum of the 21-cm fluctuations. \textsc{21CMMC} has been widely used over the past years to investigate the theoretical constrains set by different 21-cm experiments \citep{greig2015, greig2017, park2019},  break the degeneracy between different reionization topologies \citep{Binnie2019}, propose optimal designs for 21-cm instruments \citep[e.g with the SKA;][]{greig2020_skaobs}, or rule out astrophysical models incompatible with the recent upper limits obtained from LOFAR \citep{greig2020_lofar} and MWA \citep{greig2020mwa} data, respectively. In this work, we adapt it to assess the constraints set by 21-cm tomographic statistics and investigate their robustness compared to theoretical results obtained using the power spectrum. To achieve this, we create realistic mock observations using the point spread function and noise profile of the SKA1-Low telescope configuration, assuming 1000 hours of observations at redshifts 10, 9, and 8. We use \textsc{DISCCOFAN} \citep{gazagnes2019} to extract the number-count and  the morphological pattern spectra  \citep{maragos89:_patter,wilkinson2001, urbach07:_rotat_size_shape_patter_spect, westenberg2007} encoding the shape characteristics of the individual ionized regions. These statistics provide an intuitive approach to investigate the evolution of the ionizing field at different redshifts, and to infer the sources and IGM properties of the process of reionization.

This work is organized as follow: Section~\ref{sect:21cmsimu} details the implementation of \textsc{21cmFAST}, the semi-numerical code embedded in \textsc{21CMMC} that simulates the redshifted 21-cm signal during the EoR. Section~\ref{sect:morph} explains the mathematical description of the statistics extracted from the 21-cm tomographic images, and Section~\ref{sect:theory} analyzes how these statistics vary with redshift, different reionization scenarios, or are impacted by the characteristics of the interferometer. Section~\ref{sect:21CMMCsetup} details the \textsc{21CMMC} setup, and Section~\ref{sect:results} shows the results of our MCMC analysis.  Section~\ref{sect:disc} further assesses how 21-cm tomographic statistics robustly trace the underlying astrophysics of the reionization and can combined with power spectrum analyses. Finally, Section~\ref{sect:conc} summarizes our main conclusions. In this paper, we assume a $\Lambda$CDM Universe with cosmological parameters values $\Omega_\Lambda = 0.692$, $\Omega_{\rm b}$ = 0.048, $\Omega_{\rm m}$ = 0.308, $H_0$ = 67.8 km s$^{-1}$ Mpc$^{-1}$ and $\sigma_8 = 0.81$.

\section{Simulations and instrument sensitivity}
\label{sect:21cmsimu}
We use \textsc{21CMMC} \citep{greig2015}, a Bayesian parameter estimation tool which combines a  Markov Chain Monte Carlo (MCMC) algorithm with \textsc{21cmFAST} \citep{mesinger2011}, to explore how the shape of the ionized regions extracted from 21-cm tomographic images can be used to infer the properties of the sources of reionization.  Section~\ref{sect:21cmFAST} details the \textsc{21cmFAST} parametrization and Section~\ref{sect:instru} describes the interferometer characteristics adopted in this work.

\subsection{\textsc{21cmFAST}}
\label{sect:21cmFAST}
\textsc{21cmFAST}\footnote{https://github.com/andreimesinger/21cmFAST} is a publicly available semi-numerical code that simulates the evolution of the 21-cm signal during the Cosmic Dawn and the EoR using an analytical model introduced in \citet{furlanetto2004a}. Its implementation is detailed by \citet{mesinger2011}. \textsc{21cmFAST} first generates realizations of the evolved baryonic and velocity density field using the Zel'dovich approximation \citep{zeldovich1970} at different redshifts. The brightness temperature contrast, $\delta T_{\rm b}$, which is proportional to the 21-cm signal intensity and evaluated relative to the temperature of the Cosmic Microwave Background (CMB), is then computed as \citep{furlanetto2006}: 
\begin{multline}
\label{eq:21cm}
    \delta T_{\rm b} \approx 27(1-x_{\rm ion})(1+\delta_{\text{n}})\left(\frac{H(\text{z})}{\text{d}v_r/\text{d}r + H(z)}\right)\left(1-\frac{T_{\text{CMB}}}{T_S}\right) \times \\
    \left(\frac{1+z}{10} \frac{0.15}{\Omega_\text{m}h^2}\right)^{\frac{1}{2}}\left(\frac{\Omega_\text{b}h^2}{0.023}\right) \text{ mK},
\end{multline}

\noindent where $x_{\text{ion}}$ is the global ionized fraction of the IGM in the universe, $\delta_{\rm b}$ is the fractional overdensity in baryons, $H$(z) is the Hubble parameter, d$v_{r}$/d$r$ is the comoving gradient of the comoving velocity along the line of sight, $T_{\rm S}$ is the spin temperature, and $T_{\rm CMB}$ is the CMB temperature. $\delta T_{\rm b}$ is evaluated at a redshift $z$, such as $z$ = $\nu_0$/$\nu\ -\ 1$ where $\nu_0$ is the rest frame 21-cm frequency (1420 MHz). We include the effects of peculiar velocities and assume that the spin temperature is above the CMB temperature ($T_{\rm S} >> T_{\rm CMB}$) such that, in the absence of foregrounds or instrumental effects, the presence of ionized structures directly relates to the absence of 21-cm signal ($\delta T_{\rm b}$ = 0 mK). This assumption, while considered valid during most of the reionization \citep[6 $< z <$ 10,][]{chen2004, baek2010}, could break down during the early stages, or around under-dense regions where the IGM is not sufficiently heated. We discuss the impact of this hypothesis in Section~\ref{disc:limit}.

The ionized regions are recovered using an excursion set formalism that compares the number of ionizing photons per time-step to the density of baryons in regions of decreasing scales $R$ \citep{mesinger2011}. A given cell, $x$, is identified as ionized if it fulfills the following criterion:
\begin{equation}
\label{eq:ionizcrit}
   \centering \zeta f_{\rm coll}(x, z, R, M_{\rm min}) > 1,
\end{equation}
where $f_{\rm coll}(x, z, R, M_{\rm min})$ represents the collapse fraction estimated within a given scale $R$ around the cell $x$ at a redshift $z$, and depends on the minimum mass of the star-forming halo $M_{\rm min}$ \citep{press1974}. The parameter $\zeta$ characterizes the  ionizing efficiency of the sources and is further detailed in Section~\ref{sect:zeta}. \textsc{21cmFAST} additionally includes partially ionized cells by setting their ionizing fraction to the value of $ \zeta f_{\rm coll}(x, z, R, M_{\rm min})$ when $R$ reaches the minimum scale.

The focus of this study is to explore how the morphological properties of the ionized regions can be used to infer the properties of the ionizing sources. For this proof-of-concept work, we restrict ourselves to a reionization parametrization with three parameters: the ionizing efficiency of the reionization sources ($\zeta$), the mean free path of ionizing photons ($R_{\rm mfp}$), and the minimum virial temperature of dark halos hosting star-forming galaxies ($T^{\rm min}_{\rm vir}$). We detail each of them in the following sub-sections. 

\subsubsection{The ionizing efficiency $\zeta$}
\label{sect:zeta}
The ionizing efficiency of the galaxies during the EoR is defined as 
\begin{equation}
    \zeta = 30\ \frac{f_{\rm esc}}{0.3}\frac{f_{\star}}{0.05}\frac{N_\gamma}{4000}\frac{2}{1+n_{\rm rec}},
    \label{eq:zeta}
\end{equation}
where $f_{\rm esc}$ is the escape fraction of ionizing photons, $f_{\star}$  is the fraction of galactic gas in stars, $N_\gamma$ is the number of ionizing photons produced per baryon within a star, and $n_{\rm rec}$ is the typical number of recombinations of a hydrogen atom. Equation~\eqref{eq:zeta} assumes a constant ionizing efficiency for all galaxies formed in halos with sufficient mass ($M$ > $M_{\rm min}$). We note that this results in a sharp drop of the non-ionizing UV-luminosity function for halos with mass $M$ < $M_{\rm min}$ \citep[see Figure 1 in][]{greig2015}. 

Throughout, we consider a range of ionizing efficiencies from 1 to 250. As discussed in \citet{greig2017}, this  range explores a large variety of EoR models, with sources having both very low and large ionizing efficiency. The choice of $\zeta$ strongly affects the evolution of the EoR, such that increasing $\zeta$ will lead to a faster reionization process. Nevertheless, its impact also depends on the minimal virial temperature $T^{\rm min}_{\rm vir}$, since the combination of both these parameters sets the average ionizing budget during reionization.

\subsubsection{The minimum virial temperature $T^{\rm min}_{\rm vir}$}

The collapse fraction $f_{\rm coll}$ is the fraction of mass in a given volume enclosed in halos of individual mass larger or equal to $M_{\rm min}$. In \textsc{21cmFAST}, $M_{\rm min}$ is defined through the minimum virial temperature of star-forming halos, $T^{\rm min}_{\rm vir}$, and expressed as \citep{barkana2001}:
\begin{multline}
\label{eq:mmin}
    M_{\rm min} = 10^8h^{-1}\left( \frac{\mu}{0.6}\right)^{-3/2}\left(\frac{\Omega_{\rm m}\ \Delta_{\rm c}}{\Omega^{ z}_{\rm m}\ 18\pi^2}\right)^{-1/2}\left(\frac{T^{\rm min}_{\rm vir}}{1.98\ 10^4 {\rm K}}\right)^{3/2}\\
    \times \left(\frac{1+z}{10}\right)^{-3/2}M_\odot\ ,
\end{multline}
\noindent with $\mu$ the mean molecular weight, $\Omega^{z}_{\rm m}$ the evolved $\Omega_{\rm m}$ at redshift $z$ ,and $\Delta_{\rm c} = 18\pi^2 + 82d -39d$ with d =  $\Omega^{z}_{\rm m} -1$. As mentioned above, the choice of $T^{\rm min}_{\rm vir}$ impacts the position of the sharp drop in the UV luminosity function ($\zeta = 0$ if $T_{\rm vir} < T^{\rm min}_{\rm vir}$). We adopt the same prior as used in \citet{greig2017}, such that $T^{\rm min}_{\rm vir}$ varies from 10$^4$ to 10$^6$ K. The lower limit of $T_{\rm vir}$ corresponds to the minimum temperature to trigger efficient atomic cooling \citep{barkana2001, kimm2017}, while the upper limit is chosen to be consistent with the host halo mass of high redshift Lyman break galaxies observations \citep{kuhlen2012, barone2014}.

\subsubsection{The mean free path of photons $R_{\rm mfp}$}

The growth of the ionized regions is a dynamic process, which mainly depends on the recombination rate of the hydrogen atoms and the propagation distance of the ionizing photons. The balance between both processes sets the physical size of the ionized bubbles. The current \textsc{21CMMC} version explicitly computes in-homogeneous recombinations. However, we use a simplification which assumes a maximum horizon for the ionizing photons in the IGM. This parameter is denoted by $R_{\rm mfp}$ and fixes the maximum smoothing scale used in Equation~\eqref{eq:ionizcrit}. Its impact is most significant when the ionized regions size becomes closer to $R_{\rm mfp}$ \citep{greig2017}. Small values of the mean horizon of the ionizing photons (e.g. < $10$ Mpc) can delay the late stage of reionization. On the other hand, \citet{greig2017} note that values larger than 15 Mpc have little impact on the power spectrum of the 21-cm fluctuations because the fusion of the ionized regions is then the dominant growing process. We adopt a flat prior of $R_{\rm mfp} \in$ [5,25] Mpc, similar to \citet{greig2015, greig2017}.

\subsection{Interferometer characteristics}
\label{sect:instru}
We create mock observations using the point spread function (PSF) and the expected noise from 1000 hours of observations with SKA1-Low\footnote{https://astronomers.skatelescope.org/wp-content/uploads/2016/09/SKA-TEL-SKO-0000422\_02\_SKA1\_LowConfigurationCoordinates-1.pdf}. The current configuration of SKA1-Low consists of 512 stations, where 224 of them are randomly distributed in a central circular core of 500 meters in radius. The remaining 288 stations are placed in 36 clusters located in three spiral arms out to a radius of 50 km from the central core. The noise maps are computed using the python package \textsc{tools21cm}\footnote{https://github.com/sambit-giri/tools21cm} \citep{giri2020tools} which uses the formalism detailed in \citet{ghara2017} and \citet{giri2018_bsd}. The system noise per visibility is a Gaussian random variable with mean zero, and variance $\sigma^2$ defined as 
\begin{equation}
    \sigma^2 = \left( \frac{\sqrt{2}k_B T_{\rm sys}}{{\rm A_{eff}}\sqrt{\Delta \nu\Delta t}}\right)^2,
\end{equation}

\noindent where $k_B$ is the Boltzmann constant, $T_{\rm sys}$ is the telescope temperature which accounts for the receiver and sky temperature ($T_{\rm sky} \propto \nu^{-2.55}$), A$_{\rm eff}$ is the effective collective area of the individual receivers, $\Delta \nu$ is the frequency resolution of the data and $\Delta t$ is the integration time. $T_{\rm sys}$ and A$_{\rm eff}$ are fixed by the SKA1-Low technical design. Additionally, we adopt $\Delta t$ = 10 seconds per visibility, 6 hours of observing per day ($t^{\rm day}_{\rm obs}$), and a total observation time ($t^{\rm tot}_{\rm obs}$) of 1000 hours. The noise maps are obtained by generating the uv maps $U({\rm u,v,} \nu)$ using the baseline distribution in the gridded uv plane. Then, a noise cube is derived in the Fourier domain using Gaussian distributed values with mean zero and variance $\sigma^2$ for both the real and imaginary parts. The cells in the uv-plane that are not sampled are set to 0, and we divide the noise values in the remaining bins by $\frac{1}{\sqrt{U({\rm u,v,}\nu)}}$. Finally, the noise values are further scaled down by a factor $t^{\rm tot}_{\rm obs}$/$t^{\rm day}_{\rm obs}$. In reality, the baseline distribution varies in frequency such that a distinct uv map should be generated for each channel. However, in practice, these differences are small when the frequency bandwidth of the observation is relatively narrow. Consequently, we generate noise cubes using a single uv distribution which corresponds to the central frequency of the observation. Table~\ref{tab:ska} summarizes the interferometer characteristics and observation parameters for this work.

\begin{table}
	\centering
	\caption{Characteristics of the interferometer SKA1-Low and observation parameters used in this work.}
	\label{tab:ska}
	\begin{tabular}{lllllll} 
		\hline
		Parameter & Values \\
		\hline
		N$_{\rm ant}$ & 512 \\
		$T_{\rm sys}$ & 60$\left(\frac{\nu}{300 {\rm MHz}}\right)^{-2.55}$ K\\
		A$_{\rm eff}$ &  969 m$^2$ \\
		$\Delta t$ & 10 seconds \\
		$t^{\rm day}_{\rm obs}$ & 6 hours\\
		$t^{\rm tot}_{\rm obs}$ & 1000 hours \\
		\hline
	\end{tabular}
\end{table}

\section{The morphology of the ionized regions}
\label{sect:morph}
The power spectrum is a well-defined approach to analyze the excess of the 21-cm signal at different scales and already provides valuable clues about the evolution of the ionization fronts during the EoR. However, it is blind to the non-Gaussianities of the 21-cm fluctuations \citep{furlanetto2004b}, while the latter can provide important information to understand and constrain the underlying physical processes during reionization \citep{koopmans2015}. A wide range of novel approaches, using higher-order and topological statistics, have been investigated to extract the non-Gaussian information lying in the 21-cm fluctuations \citep[see Section 1 and][]{greig2019}. In this work, we define a new approach based on a pure morphological description of the ionized regions. This method provides a simplistic but intuitive description of the shape of the ionized regions, which can be efficiently extracted from noisy and PSF-convolved 21-cm image cubes. Thus, it is well suited for a Bayesian framework analysis which requires comparing thousands of models in a reasonable amount of time. Section~\ref{sect:inertia} introduces these 21-cm tomographic statistics, and Section~\ref{sect:disccoman} describes  \textsc{DISCCOFAN}, a massively parallelized image processing tool designed to extract the morphology of the structures observed in image and image-cubes.

\subsection{Morphological attributes}
\label{sect:inertia}


Several studies already highlighted theoretical differences in the morphology of the ionized regions for diverse reionization models \citep[e.g][]{kakiichi2017, giri2018_bsd, kapahtia2019, gorce2019}. In this work, we use the morphological pattern spectra \citep{maragos89:_patter}  of the ionized regions extracted from 21-cm image-cubes. This mathematical formalism is based on scale-invariant morphological attributes derived from the moment-of-inertia matrix of each ionized region. This approach has been previously introduced for detection and extraction of anomalies (aneurysms and stenoses) in blood-vessels \citep{wilkinson2001, urbach07:_rotat_size_shape_patter_spect, westenberg2007}. For a discrete case (e.g. quantized images or volumes), the moment of inertia matrix $I_o$ of a three dimensional object $O$ is defined by 
\begin{align}
\begin{split}
I_o &=
\begin{bmatrix}
    I_o^{xx}   & I_o^{xy} & I_o^{xz} \\
    I_o^{yx}   & I_o^{yy} & I_o^{yz} \\
    I_o^{zx}   & I_o^{zy} & I_o^{zz}
\end{bmatrix} \quad \text{with} \ \\
 I_o^{xx} & = \sum_O(x-\bar x)^2 + V_o/12 \\
I_o^{yy} & = \sum_O(y-\bar y)^2 + V_o/12\\
I_o^{zz} & = \sum_O(z-\bar z)^2 + V_o/12\\
I_o^{xy} & = \sum_O(x-\bar x)\times(y-\bar y) \\
I_o^{xz} & = \sum_O(x-\bar x)\times(z-\bar z), 
\end{split}
\end{align}

\noindent where [$x$, $y$, $z$] are the coordinates of the voxels belonging to $O$, [$\bar{x}$, $\bar{y}$, $\bar{z}$] are the coordinates of its center of mass, and $V_o$ is its volume. This formalism is derived from the continuous case where $I_o^{xx} = \iiint_O x^2\, dV\ $. The factor $V_o/12$ is introduced to account for the moment of inertia of the individual cubic voxels. From $I_o$ and its eigenvalue decomposition, we extract four moments-invariant attributes encoding different morphological properties of a three dimensional geometrical structure: the \textit{Elongation} $\mathcal{E}$, the \textit{Flatness} $\mathcal{F}$, the \textit{Non-compactness} $\mathcal{N}$, and the \textit{Sparseness} $\mathcal{S}$.

\paragraph*{\textit{Non-compactness}:} $\mathcal{N}$ is defined as 

\begin{equation}
    \mathcal{N} = \frac{{\rm Tr}(I_o)}{V_o^{5/3}},
\end{equation}

\noindent where ${\rm Tr}(I_o)$ is the trace of the moment of inertia tensor $I_o$. The unit of $I_o$ is consistent with a length L to the power 5 while the volume scales with L$^3$, hence the ratio $\frac{{\rm Tr}(I_o)}{V_o^{5/3}}$ is scale-invariant. $\mathcal{N}$ reaches a minimum value of 0.25 for perfectly spherical objects and increases for complex asymmetric structures. 

\paragraph*{\textit{Elongation} and \textit{Flatness}:} $\mathcal{E}$ and $\mathcal{F}$ are derived using the ratio of the eigenvalues of $I_o$, referred as $\lambda_1$, $\lambda_2$ and $\lambda_3$ with $|\lambda_1| \geq |\lambda_2| \geq |\lambda_3|$, such that

\begin{align}
\begin{split}
    \mathcal{E} &= \frac{|\lambda_1|}{|\lambda_2|} \quad {\rm and} \\
    \mathcal{F} &= \frac{|\lambda_2|}{|\lambda_3|}.
\end{split}
\end{align}

\noindent The eigenvalues $\lambda$ represent the variance of the coordinates of an object along its main axes, scaled by its volume. In other words, they probe the spatial expansion of a three dimensional structure along the directions of maximal growth, such that $|\lambda_1| = |\lambda_2| = |\lambda_3|$ for perfectly spherical objects or $|\lambda_1| > |\lambda_2| = |\lambda_3|$ for elongated shapes. Thus, $\mathcal{E}$ and $\mathcal{F}$ refer to the divergence between the object growth in different directions. Combined, they provide information about the eccentricity of a geometrical shape. From the description above, it comes trivially that spherical objects have $\mathcal{E} = \mathcal{F} = 1$, cylindrical or cigar-shaped objects have $\mathcal{E} > 1$ and $\mathcal{F} \approx 1$, and more complex tri-axial shapes have $\mathcal{E}$ and $\mathcal{F}$ larger than 1. The analysis of $\mathcal{E}$ and $\mathcal{F}$ is useful to reveal a dominant direction in the growth of the ionized regions, and therefore provide information whether the propagation of the ionizing radiation is isotropic. 

\paragraph*{\textit{Sparseness}:} $\mathcal{S}$ is defined as

\begin{equation}
     \mathcal{S} = \frac{\pi 20^{3/2} \textstyle \prod\limits_{i}^3 { |\lambda_i|}}{6V_0^{5/2}}.
\end{equation}

\noindent It corresponds to the ratio of the expected volume computed using the coordinate variance along the three main axes (product of the three eigenvalues), versus the actual volume, obtained by summing over all the voxels belonging to the object. Broadly speaking, $\mathcal{S}$ relates to the filling factor of an object. It reaches 1 for filled spheres or ellipsoids, 1.127 for solid rectangular blocks, and increases as the structures become more porous.

Overall, these four morphological attributes provide different information about the shape of the ionized regions.  $\mathcal{E}$ and $\mathcal{F}$ highlight  the structures eccentricity, while $\mathcal{N}$ and $\mathcal{S}$ describe the shape complexity, through its degree of asymmetry and porosity. Throughout this work, we also consider the total number of individual ionized regions observed, $n_{\rm bub}$, and the volume of the ionized regions, V. Both measures are reliable probes of the underlying astrophysics of reionization \citep[e.g.][]{kakiichi2017, giri2018_bsd}. In the next section, we detail the algorithm used to extract $n_{\rm bub}$, the volume, and four morphological attributes of each ionized region directly from the three-dimensional 21-cm images.

\subsection{Extracting morphological attributes with DISCCOFAN}
\label{sect:disccoman}

To extract the number and the morphological pattern spectra of the individual ionized regions from 21-cm tomographic images, we use \textsc{DISCCOFAN}\footnote{\url{https://github.com/sgazagnes/DCF-2D}} \citep[DIStributed Connected COmponent Filtering and ANalysis;][Gazagnes \& Wilkinson, submitted]{gazagnes2019}, a recent method that combines image processing and mathematical morphology techniques to efficiently find, select, and analyze the connected structures in two and three dimensional data sets. Contrary to classical pixel-based approaches, \textsc{DISCCOFAN} works on a region-based representation of the data referred to as a \textit{component tree} \citep{salembier98}. During the first step (the \textit{flooding}), the algorithm goes through all the voxels and group them in \textit{connected components}. The typical clustering rule to define a connected component is to select all the neighbors' voxels above a given threshold set. The nested relations between the connected components at different intensity levels are stored in a tree structure, which allows us to perform  complex multi-scale operations on this optimized hierarchical representation. The use of similar tree-techniques has been successfully used to improve the detection of faint-objects in optical astronomical surveys \citep{teeninga16, haigh2020}. 

The procedure to derive the ionized region statistics from noisy, low-resolution 21-cm image cubes remains the same for all the tests performed in this work. We first extract the ionized regions directly from the 21-cm mock observations using an independent thresholding technique (see more details in Section~\ref{sect:noisepsf}). \textsc{DISCCOFAN} is then applied on the resulting segmented image cubes encoding the recovered ionized regions such that voxels with a value of 0 and 1 are defined as neutral and ionized, respectively. \textsc{DISCCOFAN} finds and extracts all the individual connected components that have spatially connected voxels with a value equal to 1, assuming that each voxel has 26 neighbors in three dimensions. The total number of connected components found corresponds to the number of ionized regions $n_{\rm bub}$. The volume of each region is derived by summing over all the voxels that belong to it, and by multiplying this number with the volume of the voxel. Finally, \textsc{DISCCOFAN} simultaneously returns the morphological attributes $\mathcal{E}$, $\mathcal{F}$, $\mathcal{N}$, and $\mathcal{S}$ using the equations presented in Section~\ref{sect:morph}. As applied on binary image cubes, the approach of \textsc{DISCCOFAN} is very similar to a classical friend-of-friend algorithm. However, this tool provides a powerful framework to analyze the full range of 21-cm intensities, and we plan to explore this aspect in future works.

\textsc{DISCCOFAN} can process very large  data-sets ($> 100$ Gigapixels; Gazagnes \& Wilkinson, submitted). It is massively parallelized using shared and distributed memory techniques, thus is fast and efficient to analyze the properties of connected structures in any data set. Using a single CPU process, it takes around 10 seconds to create a model on a 128$^3$ box with \textsc{21cmFAST}, while it takes less than a second to extract the morphological pattern spectra with \textsc{DISCCOFAN}. This asset makes this tool well suited to be incorporated in a Bayesian inferential framework, where thousand of models need to be processed in a reasonable time.

\section{Morphology of the ionized regions }
\label{sect:theory}
In this section, we investigate the evolution of the morphological pattern spectra of the ionized regions as a function of redshift (Section~\ref{sect:evolution}), different reionization scenarios (Section~\ref{sect:model}), and the impact of the expected noise profile and PSF of SKA1-Low on the recovered statistics (Section~\ref{sect:noisepsf}).

\subsection{Evolution during reionization}
\label{sect:evolution}

\begin{figure*} 
\includegraphics[width=\textwidth]{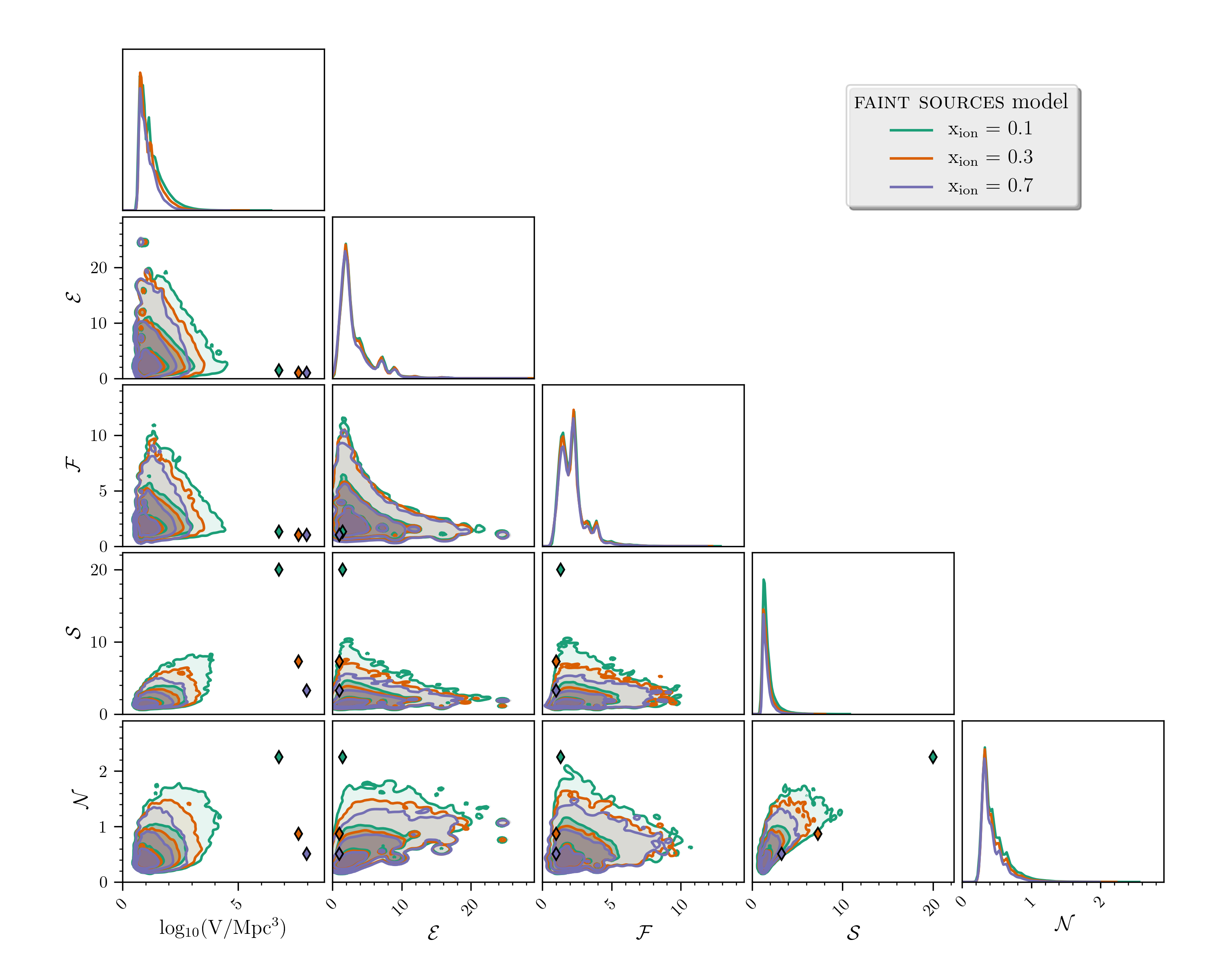}
 \caption{Evolution of the morphological attributes of the  ionized regions for the \textsc{faint sources} model simulations (512$^3$ Mpc$^3$), for three snapshots corresponding to an ionized fraction $x_{\rm ion}$ of 0.1, 0.3 and 0.7, respectively. The diagonal and lower panels show respectively the 1D and 2D marginalized  PDFs. The contours enclose successively 68, 95, and 99.7~\% of the ionized region distributions. The diamonds show the morphological attribute values of the largest ionized region in each cube.}
 \label{fig:hiimock}
\end{figure*}

We first investigate the evolution of the morphological attributes (V, $\mathcal{E}$, $\mathcal{F}$, $\mathcal{S}$, and $\mathcal{N}$)  during reionization using simulated 21-cm maps. We generate co-eval boxes of 512$^3$ cells with 2 Mpc resolution using \textsc{21cmFAST} with the set of parameters $\zeta = 30$,  $T^{\rm min}_{\rm vir} = 5.10^4 K$, and $R_{\rm mfp} = 15$ Mpc (referred as the \textsc{faint sources} model). This model characterizes a cosmic reionization dominated by a population of several sources with a relatively low ionizing efficiency \citep{mesinger2016, greig2017}. We consider three different redshifts corresponding to snapshots where the ionizing fraction ($x_{\rm ion}$) in the box is successively 0.1, 0.3 and 0.7. This choice enables us to investigate the evolution of the ionized regions morphology for three different reionization stages where the universe is mostly dominated by (1) individual ionized regions; (2) ionized filaments resulting from the fusion of the ionized structures; and (3) neutral islands spanning through a main ionized percolated region which fills most of the universe \citep{chen2019}. For this noise-free exploratory case, we extract the ionized regions from the synthetic 21-cm maps using $\delta T_{\rm b} = 0$ mK as a threshold, and we apply \textsc{DISCCOFAN} on the resulting three-dimensional binary images.

Figure~\ref{fig:hiimock}\footnote{Corner plots from Figures~\ref{fig:hiimock}, \ref{fig:hiimodel} and \ref{fig:hiinoise} have been made using the python package \textsc{pappy}: \url{https://github.com/drphilmarshall/pappy}.} shows the resulting distributions of the five morphological attributes for the three snapshots as a corner plot. The diagonal and lower panels show respectively the one and two-dimensional marginalized  probability density functions (PDFs). The contours in the lower panels enclose successively 68, 95, and 99.7~\% of the ionized regions distributions. Additionally, we highlight the morphological properties of the largest ionized region in each box using diamonds. The largest scales are typically more sensitive to sample variance. However, resolving the properties of individual ionized regions with SKA is expected to be simpler as their size grows larger \citep{ghara2019}. Hence, it should be possible to extract valuable information about the astrophysics of the reionization process using the morphological properties of these large structures.

We observe that the typical size of the ionized regions shrinks as reionization progresses, while the largest objects grow larger. The probability to observe several large ionized structures decreases because more and more of these regions fuse into a larger \textit{percolated} cluster, spanning the whole box,  with boundaries that are artificially infinite. Figure~\ref{fig:hiimock} shows that a large ionized regions of $\approx 10^7$ Mpc$^3$ is already present at $x_{\rm ion} = 0.1$, suggesting that a percolated object already formed at this early stage. This is consistent with \citet{furlanetto2016} who found that the formation of a percolated cluster typically happens when the universe is still only mildly ionized ($x_{\rm ion}$ between 0.1 and 0.2). 

The distributions of $\mathcal{E}$ and $\mathcal{F}$ also provide some interesting insights about the growth of the ionized regions. The elongation and flatness are large when the ionized regions are small, but decrease as their volume gets larger, with $\mathcal{E}$ and $\mathcal{F}$ tending closer to 1. This suggests that the growth mechanisms of the small structures are likely anisotropic, but the fusion of the individual regions equalizes the spatial extension of the larger merged structures. However, the $\mathcal{E}$ and $\mathcal{F}$ values derived in the largest  ionized region should be taken with caution. Indeed, the box has a finite length, such that the maximal spatial extension of an object is limited by the size of the simulation. Hence, the percolated cluster  will always have  $\mathcal{E} \approx \mathcal{F} \approx 1$, because it spans over the whole box. 

Interestingly, the two dimensional PDF of $\mathcal{F}$ versus $\mathcal{E}$ highlights that only few objects have $\mathcal{E}$ and $\mathcal{F}$ = 1 which supports that the shape of the ionized structures strongly diverges from the idea of spherical bubbles. On the other hand, objects with the largest values of $\mathcal{E}$  ($\mathcal{F}$) have $\mathcal{F}$ ($\mathcal{E}$) closer to 1, suggesting that these regions are more extended along a specific direction, and resemble cylindrical or filamentary structures. This outcome is somewhat expected from \textit{inside-out} reionization,  as implemented in \textsc{21cmFAST} \citep[Furlanetto, Hernquist, and Zaldarriagan (FZH) model;][]{furlanetto2004a}. In such a model, the star formation is predominantly located on the high-density filaments such that these regions are ionized first. Thus, the ionized regions  grow and merge along with these filamentary structures, which likely explains the observed shape. Additionally, the distributions of $\mathcal{E}$ and $\mathcal{F}$ remain fairly constant as reionization progresses, which might indicate that these quantities probe independent IGM properties, such as the structure of the underlying density field. Recent studies have shown that the power spectrum should be able to provide information about the topological property of the reionization (\textit{outside-in} or \textit{inside-out}) because it can indirectly probe the correlation between the ionization and density field  with the high-density regions \citep{Binnie2019, pagano2020}. 

Finally, Figure~\ref{fig:hiimock} highlights that larger objects are typically more porous and less compact than smaller ionized regions. This is somewhat expected since the formation of the largest objects is mostly driven by the fusion of individual ionized regions, such that the resulting structures are more complex and porous.  The values of $\mathcal{S}$ and $\mathcal{N}$ decrease as reionization evolves because the neutral regions within the ionized filaments become progressively ionized. This can be seen as the 99.7\% contours enclose a narrower range of $\mathcal{S}$ and $\mathcal{N}$ for ionized regions with the same volume at any redshift. This effect is also particularly noticeable on the properties of the percolated cluster in each box. As reionization progresses, these objects become more compact and ``filled" because the remaining neutral patches are progressively ionized. Hence, even when the universe is predominantly ionized, comparing the morphology of these percolated clusters might still provide valuable insights about the properties of the ionizing sources. This is further discussed in the next section.

Overall, Figure~\ref{fig:hiimock} emphasizes that the evolution of the morphology of the ionized regions probes the reionization history. Mainly, early reionization stages form complex, eccentric, and overall porous ionized structures due to the fusion of the individual ionized regions along the high-density filaments of the underlying density field. As reionization progresses, these structures continue to grow as their ionizing fronts are propagating. However, their maximal size becomes limited by the merging mechanisms, such that they are more likely to fuse early with the percolated cluster. In the next section, we compare these statistics for two different reionization scenarios.

\subsection{Morphological spectra for different reionization scenarios}
\label{sect:model}

\begin{figure*} 
\includegraphics[width=\textwidth]{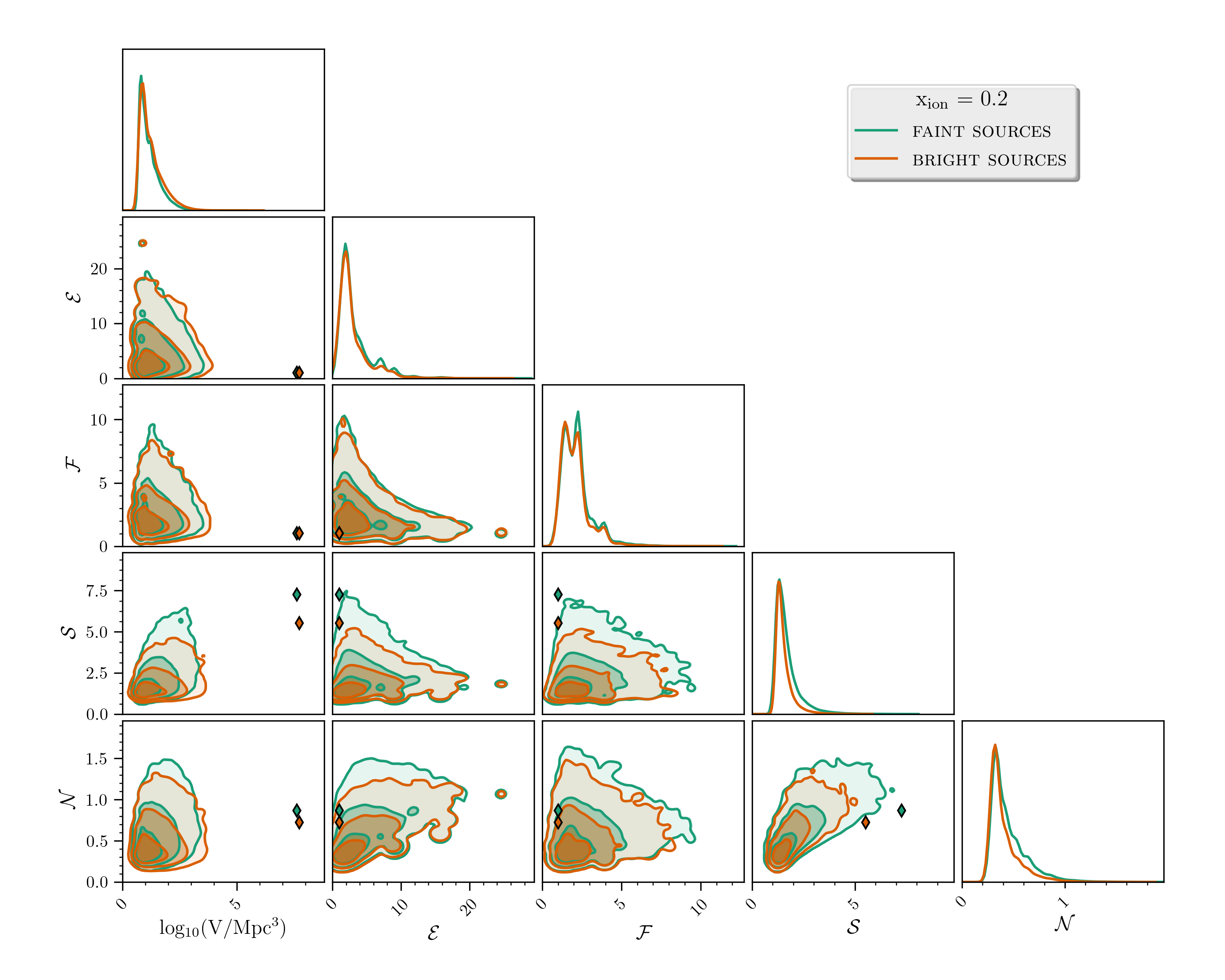}
 \caption{Comparison of the observed distribution of the ionized regions morphological attributes for the \textsc{faint sources} and \textsc{bright sources} models. The morphological properties are extracted on a snapshot where the global ionized fraction is 0.2. The contours enclose successively 68, 95, and 99.7~\% of the ionized regions distributions. The diamonds show the morphological attribute values of the largest ionized region in each cube.}
 \label{fig:hiimodel}
\end{figure*}

Understanding how the neutral hydrogen was ionized during to reionization requires one to constrain the properties of the sources that dominantly contributed to the ionizing budget of the EoR. Studies generally investigate the impact of different populations of reionization sources, labeled through their ionizing efficiency, the typical mass of the host halos, or the hardness of their X-ray spectral energy distribution (SED). We use the \textsc{faint sources} and \textsc{bright sources} models, previously introduced in \citet{mesinger2016} and \cite{greig2017}, and defined by the following sets of parameters:
\begin{itemize}
    \item \textsc{faint sources}:  $\zeta = 30$,  $T^{\rm min}_{\rm vir} = 5.0\times 10^4 K$, and $R_{\rm mfp} = 15$ Mpc.
    \item  \textsc{bright sources}:  $\zeta = 200$,  $T^{\rm min}_{\rm vir} = 3.0\times 10^5 K$, and $R_{\rm mfp} = 15$ Mpc.
\end{itemize}

\noindent The \textsc{faint sources} model, already introduced in Section~\ref{sect:evolution}, characterizes a reionization history dominated by a large population of galaxies with a relatively low ionizing efficiency. On the other hand, the \textsc{bright sources} model has larger $\zeta$ and  $T^{\rm min}_{\rm vir}$, and is characterized by fewer sources with larger ionizing efficiency. By construction, these models have fairly similar reionization histories \citep[see Figure 2 in][]{greig2017} and  match both the inferred evolution of the cosmic star formation rate (SFR) density using the extrapolated observed luminosity functions of \citet{bouwens2015} and the electron scattering optical depth, $\tau_e$  = 0.058 $\pm$ 0.012 from \citet{planck2016}. These two scenarios have strong implications for the patchiness of reionization and the morphology of the observed ionized regions, and provide a first glance at what differences would we observe for an EoR driven by AGNs or by fainter galaxies \citep{robertson2015, finkelstein2019}. We note that these models are not unique to explain reionization, as recent studies have shown that ``intermediate" scenarios, driven by brighter galaxies, could very well match the current observational constraints (e.g Lyman $\alpha$ damping wings of quasars and galaxies at $z$ > 7) \citep{naidu2020}. Nevertheless, they are interesting for astrophysical parameter forecasting frameworks to explore which information is needed to favor or rule out specific scenarios. Similar to Section~\ref{sect:evolution}, in this noise-free case, we extract the ionized regions from the synthetic 21-cm maps using $\delta T_{\rm b} = 0$ mK as the threshold value.

In Figure~\ref{fig:hiimodel}, we investigate the distribution of the morphological attributes of the ionized regions for these two models using a snapshot with $x_{\rm ion} \approx 0.2$. This ionizing fraction corresponds to  redshift 9 and 8.5 for the \textsc{faint sources} and \textsc{bright sources}, respectively.  Similarly to Figure~\ref{fig:hiimock}, the contours plotted in the lower panels include successively 68, 95, and 99.7 \% of the distributions and we highlight the properties of the largest (percolated) ionized region in the box with diamonds. 

We note several modest differences between the observed distributions for both models. The size distribution of the ionized regions suggests that larger ionized structures have formed when reionization is dominated by brighter sources. Additionally, the distributions of $\mathcal{S}$ and $\mathcal{N}$ show that the \textsc{faint sources} model produces more sparse and asymmetric ionized regions. We note that these differences are more significant in the tails of these distributions, suggesting that outliers are important to discriminate between these different reionization scenarios.

The distributions of $\mathcal{E}$ and $\mathcal{F}$ stay roughly similar for both models. As mentioned above, this might suggest that this property is independent of the reionization scenario as it probes the correlation between the ionization field and the underlying density field. This property should prove useful to investigate the global topology of reionization or more complex reionization models, as discussed in Section~\ref{sect:evolution}. 

The morphology of the percolated object in each model  provides similar insights. It appears more porous and less compact in the \textsc{faint sources} model. This is expected since a population of fainter sources form many individual ionized regions, whose fusions result in more porous and larger structures compared to a reionization scenario driven by fewer brighter sources. Interestingly, both percolated clusters have similar size, suggesting that their morphological properties can provide important additional insights on the nature of the underlying sources. 

We note that, in this section and Section~\ref{sect:evolution}, we used a single set of initial conditions to perform this analysis. In theory, the tails of the different distributions presented in Figures~\ref{fig:hiimock} and \ref{fig:hiimodel} might change for different set of initial conditions. This cosmic variance effect can have significant consequences when comparing the observed ionized regions number-count and morphology in various astrophysical models. Nevertheless, we found that, when using large box sizes (512$^3$ cells with 2 Mpc resolution), the impact of cosmic variance is negligible compared to the difference between these two models. Additionally, in Section~\ref{sect:sample}, we further assess this effect on our inferential framework using smaller mock observations of 128$^3$ cells with a 250 Mpc box size. We show that using different sets of initial conditions does not significantly impact the recovered parameter intervals, which suggests that the variations due to cosmic variance is typically smaller than the variation between astrophysical models.

Overall, this section highlights that the most significant differences between these two reionization scenarios are related to the size, sparseness, and compactness of the ionized regions. Additionally, while we did not discuss it in this section, the total number of observed ionized regions should also provide additional information to compare different models. Several studies showed that we should observe fewer ionized regions for scenarios with brighter sources, or with a higher minimum mass of the star-forming halos \citep{kakiichi2017,giri2018_bsd}. Nevertheless, these differences are more complicated to extract from realistic observations, as instrumental effects tend to smooth the contours of the ionized regions and suppress the information lying in the smaller scales.

\subsection{Impact of the point spread function and noise}
\label{sect:noisepsf}

\begin{figure*} 
\includegraphics[width=\textwidth]{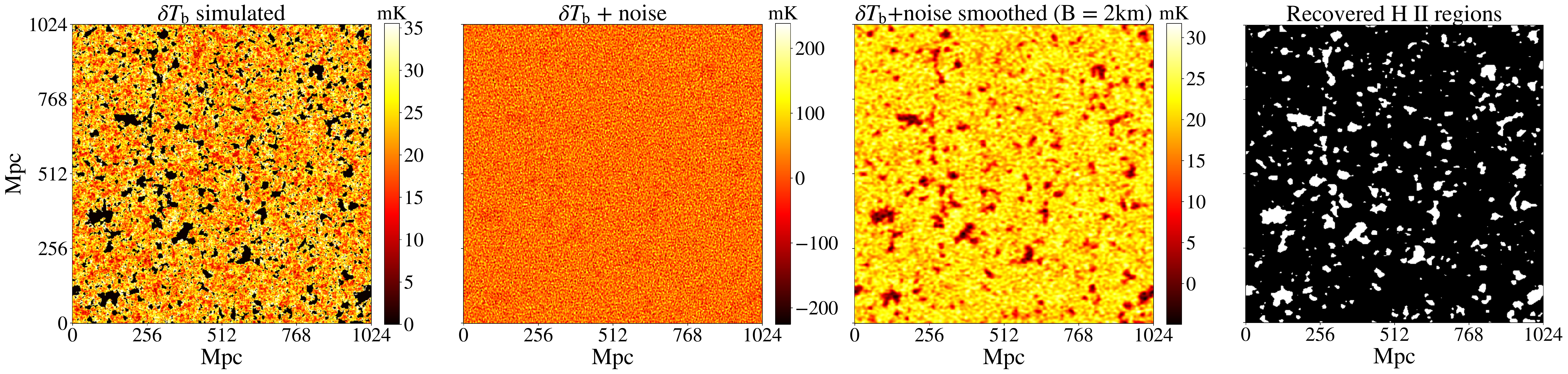}
\caption{Two dimensional slices extracted from the 512$^3$ cells co-eval 21-cm signal cube at $z$ $\approx$ 9 ($x_{\rm ion}$ = 0.20)  for the \textsc{faint sources} model. From left to right: the simulated brightness temperature intensity map with an intrinsic resolution of 2 Mpc (0.730 arcmin) resolution; the 21-cm signal observed with noise corresponding to the maximum baseline of 10 km with SKA1-Low; the 21-cm signal observed after applying a smoothing kernel in space and frequency corresponding to a maximum baseline of 2 km ($\approx$ 2.9 arcmin  or 7.9 Mpc resolution); the extracted ionized regions using the \textit{Triangle} thresholding approach (see details in Section~\ref{sect:noisepsf}).}
 \label{fig:mockimages} 
\end{figure*}

\begin{figure*} 
\includegraphics[width=\textwidth]{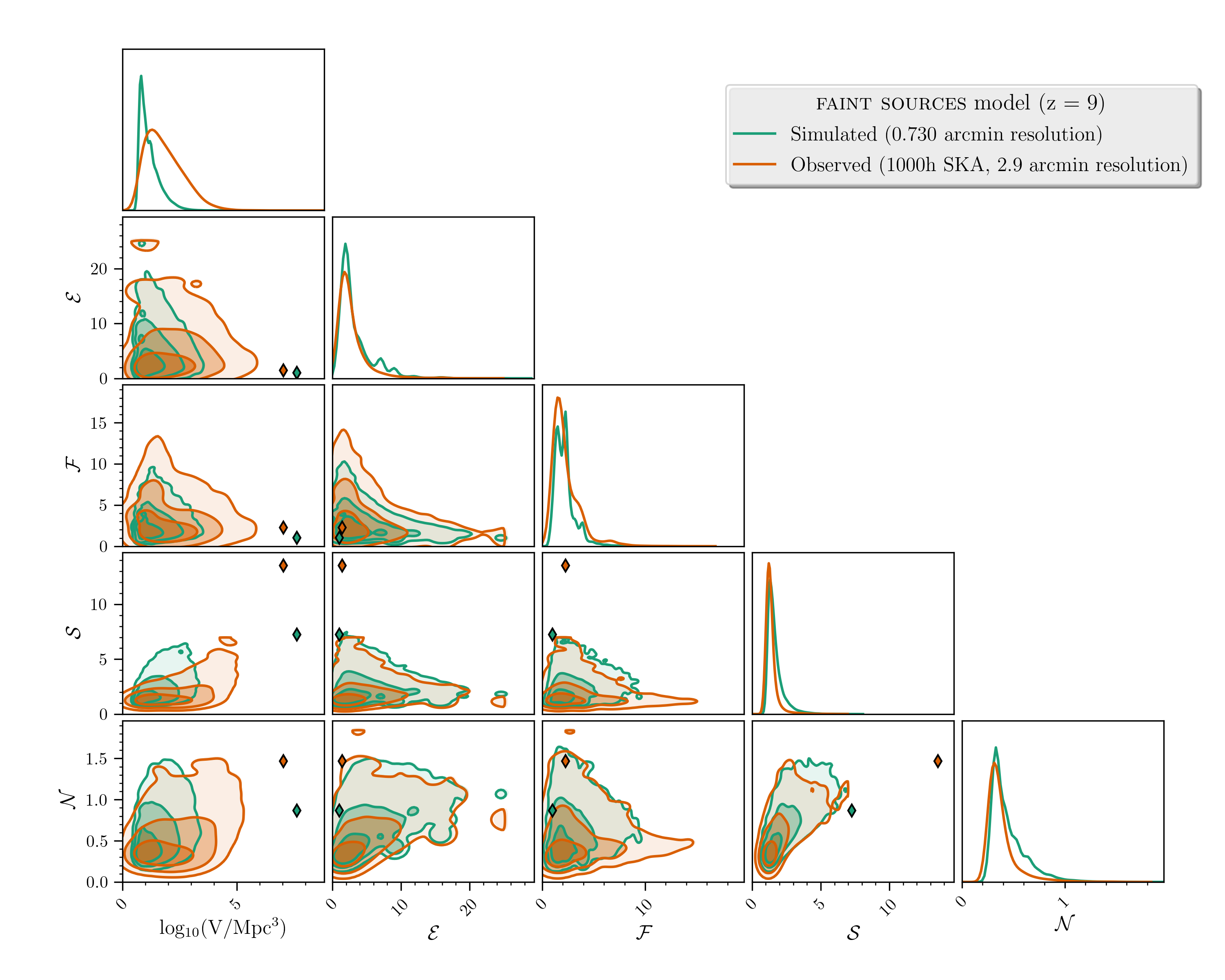}
 \caption{Comparison of the morphological attributes of the ionized regions recovered on the synthetic simulation (green) and on the realistic mock observation assuming 1000 hours observations with SKA1-Low (orange), using a snapshot of the \textsc{faint sources} model with $x_{\rm ion}$ = 0.20. The contours enclose successively 68, 95, and 99.7~\% of the ionized regions distributions. The diamonds show the morphological attribute values of the largest ionized region in each cube.} 
 \label{fig:hiinoise}
\end{figure*}

In this section, we investigate the robustness of the morphological properties of the ionized regions extracted from realistic 21-cm tomographic observations with SKA1-Low. To do so, we simulate noise cubes using the package \textsc{tools21cm} (which follows the procedure defined in Section~\ref{sect:instru}), and add them to the simulated $\delta T_{\rm b}$ boxes. Because the root mean square (rms) of the noise is too large compared to the 21-cm signal, we additionally smooth the data by applying a Gaussian smoothing in the spatial direction and a top-hat filter in the frequency direction such that the full width at half maximum corresponds to a maximum baseline of 2 km  \citep[similarly to][]{giri2018_bsd, giri2018_identif}. This is because most of SKA's collecting area is at baselines smaller than 2 km. This additional   smoothing increases the S/N of the observation, but also smooths out the contours of the ionized regions. Throughout, we assume perfect foreground removal, but we investigate the impact of potential foreground residuals in Section~\ref{sect:gauss}.

Instrumental effects complicates the extraction of the ionized regions from 21-cm tomographic observations. To optimally identify the ionized regions, \citet{giri2018_identif} recently proposed a new approach using a super-pixel thresholding method \citep[SLIC;][]{achanta2012}. The authors show that this method performs better than typical segmentation approaches when the ionization fraction of the box is larger than 0.1. Nevertheless, this method is relatively time-consuming, making it less attractive when including it  within a Bayesian inference framework such as \textsc{21CMMC}. We choose a different thresholding approach, referred to as the \textit{Triangle thresholding} \citep{zack1977}. This strategy is based on the PDF of the pixel intensities in the image. It finds the optimal threshold value by constructing a line between the histogram peak and the farthest measurement in the tail of the histogram.  The threshold is derived by taking the point of maximum distance between this line and the histogram level. This technique is also referred to as the maximum deviation method in \citet{giri2018_identif}, and is found to be particularly effective for bi-modal distributions. Hence, it is well suited to extract the ionized regions from noisy 21-cm observations because the presence of these structures should imprint a clear peak on the PDF of the 21-cm intensities for observations well within the EoR. Nevertheless, this might be more complex for the very early stages where the ionizing fraction is too low to robustly identify these regions \citep{kakiichi2017, giri2018_identif}. Overall, we found that this method was giving good performance up to $x_{\rm ion}$ > 0.05, and that it gives similar performance as the SLIC approach (see Appendix~\ref{app:thresh}). The choice of our particular approach is motivated by its better computational efficiency. Using a single CPU process and a mock 21-cm image cube of 128$^3$ cells, it takes less than 1 second to derive the segmented image using the \textit{Triangle thresholding} approach, while SLIC takes more than 10 seconds.

To investigate the effects of noise and resolution on the ionized region morphology, we compare the results obtained in Section~\ref{sect:evolution} for the \textsc{faint sources} model using a snapshot where $x_{\rm ion} = 0.2$. We display on Figure~\ref{fig:mockimages} the impact of noise on the simulated three dimensional image cubes, by showing from left to right the two dimensional slices corresponding to the simulated 21-cm image cube; after adding the noise; after smoothing using a kernel corresponding to a maximum baseline of 2 km; and after extracting the ionized regions using the \textit{Triangle thresholding} applied on the noisy low resolution 21-cm tomographic observations. The ionized regions statistics are derived by applying \textsc{DISCCOFAN} (Section~\ref{sect:disccoman}) on the latter segmented image cube.

In Figure.~\ref{fig:hiinoise}, we compare the recovered distributions of the ionized regions morphological attributes in the noise-free case and after including the noise and smoothing, using the same corner plots as in Figures~\ref{fig:hiimock} and \ref{fig:hiimodel}. Typically, instrumental effects tend to smooth the observed distributions of the morphological properties. We note that this effect is more significant for the size distribution of the ionized regions, which peaks at larger object sizes. This is expected from previous studies which showed that smoothing strongly affects the observed distribution of object sizes, because it artificially merges ionized regions that are originally disconnected, and thus indirectly  decreases the number of individual ionized regions observed \citep{kakiichi2017, giri2018_bsd}.

The contours of the 2D PDF of $\mathcal{E}$ and $\mathcal{F}$ are slightly larger, suggesting that they appear more eccentric than in the simulated maps. We note that this might be because we use a different smoothing procedure in the spatial and frequency space, with respectively a Gaussian and top-hat kernel, which could accentuate their extension along specific directions. 

Additionally, the ionized regions extracted from realistic observations are overall less porous and more compact, especially regarding the smallest structures, since the PSF of the instrument, and the additional 3D smoothing will smooth the contours and the shapes of the ionized regions. 

Interestingly, the percolated cluster has a smaller volume, but larger values of  $\mathcal{E}$, $\mathcal{F}$, $\mathcal{S}$ and $\mathcal{N}$, suggesting that the cumulative effects of noise and smoothing increase the peculiar morphology of this large structure. This is likely because, at this stage, the percolated cluster is already very sparse, such that the instrumental effects  accentuate its porosity and overall complexity. 

While the size distribution is slightly shifted towards larger objects, Figure.~\ref{fig:hiinoise} shows that the four other shape statistics are fairly accurately recovered on noisy low-resolution observations. We note however that the impact of smoothing strongly depends on the reionization stage, and on the contrast of the 21-cm signal fluctuations which can vary for different reionization scenarios. Consequently, accurately analyzing the impact of the instrumental effects of the recovered ionized morphology should be done as a function of multiple reionization stages and different reionization scenarios. However, the performance of the inference framework  will indirectly reveal whether the extracted statistics are robust enough against these effects.



\section{21CMMC setup}
\label{sect:21CMMCsetup}

\textsc{21CMMC}\footnote{https://github.com/BradGreig/21CMMC} is a MCMC sampler designed to explore the astrophysical parameter space of the Cosmic Dawn (also sometimes called the Epoch of Heating; EoH) and EoR. It includes a modified version of the python module \textsc{Cosmohammer} \citep{akeret2013} which is built on the top of the \textsc{emcee} python module \citep{foreman2013} using an affine invariant ensemble sampler \citep{goodman2010}. \textsc{21CMMC} employs a streamlined version of \textsc{21cmFAST} (see Section~\ref{sect:21cmFAST}) to efficiently simulate the 21-cm brightness temperate fluctuations for different sets of astrophysical parameters. It has been designed to prepare forthcoming observations and already used in many studies to quantify the constraints and degeneracies among the reionization model astrophysical parameters  \citep[e.g.][]{greig2015, greig2017, greig2018, park2019}. In this work, we use it to explore the use of 21-cm tomographic statistics as an alternative to the power spectrum to recover the EoR astrophysical parameters. In Section~\ref{sect:likeps}, we summarize the original implementation of \textsc{21CMMC} based on the power spectrum of the 21-cm observations. We detail, in Section~\ref{sect:likemorph}, the new likelihood function implemented in \textsc{21CMMC}, which uses the ionized regions statistics extracted from 21-cm images. Finally, in Section~\ref{sect:mock},  we present the mock observations and general setup used in this work. 


\subsection{21CMMC using the power spectrum}
\label{sect:likeps}

In \textsc{21CMMC}, the likelihood function for the 21-cm power spectrum ($\Delta$) is defined as a $\chi^2$ statistics. It is expressed as

\begin{equation}
\label{eq:likeps}
    \mathcal{L}(\Delta) =  \sum_{i = 1}^{N_k}{\exp \left(\frac{\left(\Delta_{\rm obs}(k_i)^2 - \Delta_{ \rm mod}(k_i)^2\right)^2}{2\sigma_\Delta(k_i)^2}\right)} 
\end{equation}
where $\Delta_{\rm obs}$ and $\Delta_{\rm mod}$ are the power spectrum of the observation and the model, respectively, $k$ is the Fourier mode, $N_k$ the number of independent Fourier modes included (8 in this work), and $\sigma_{\Delta}$ is the 21-cm power spectrum uncertainty defined as

\begin{equation}
    \sigma_{\Delta}(k_i)^2 = \sigma^{\rm thermal}(k_i)^2 + \sigma^{\rm variance}_{\rm obs}(k_i)^2 + \sigma^{\rm variance}_{\rm mod}(k_i)^2,
\end{equation}

\noindent where $\sigma^{\rm thermal}$ is the thermal noise, and $\sigma^{\rm variance}_{\rm obs}$ and $\sigma^{\rm variance}_{\rm mod}$ are the sample variance of the observation and model, respectively. In this work, $\sigma^{\rm thermal}$ is estimated by deriving the uv coverage for a 1000h observation with SKA1-Low (see parameters in Table~\ref{tab:ska}), simulating the thermal noise using a SEFD of 2500 Jy at the central frequency of the observation, and generating the corresponding thermal noise power spectrum and uncertainty\footnote{https://gitlab.com/flomertens/ps\_eor}.  We note that, in \citet{greig2015, greig2017}, the authors included an additional  \textit{modelling} uncertainty, typically fixed to 20\% of the mock power spectrum value, to account for the semi-numerical approximations compared to fully numerical codes \citep{zahn2011,hutter2018}, and differences in the radiative transfer equations implementation \citep{iliev2006}. Nevertheless, in this work, we aim to investigate and compare the performance of a parameter inferential approach using the ionized regions morphological pattern spectra and using the 21-cm power spectrum. Hence, including such error term for the power spectrum would require an equivalent for the 21-cm tomographic statistics, which is not trivial to define. Consequently, we choose to exclude this modelling uncertainty for both cases. Additionally, the 8 independent $k$ bins are sampled in logarithmic space in the interval [0.1, 1] Mpc$^{-1}$. The lower bound of this interval corresponds to the foreground corruption limit while the upper bound is fixed by the thermal-noise limit.

\subsection{21CMMC using tomographic statistics}
\label{sect:likemorph}


Defining a likelihood function to compare the distribution of the morphological attributes of the ionized regions is not trivial. We need to compare two distributions of $n_{\rm bub}$, 5-dimensional vectors $\vec{v}$, such that $\vec{v}$ = [V, $\mathcal{E}$, $\mathcal{F}$, $\mathcal{S}$, $\mathcal{N}$], and  $n_{\rm bub}$ is the number of ionized regions extracted in the 21-cm images\footnote{In practice, we only keep the ionized regions that have a volume larger than 10 Mpc$^3$. This is because lower scales are heavily affected by the noise and the resolution of the instrument.
}. Comparing 5-dimensional distributions can be computationally costly and penalize the MCMC run if the time taken to compute the likelihood value is too large with respect to the time required to simulate the models. Additionally, the problem is complex because $n_{\rm bub}$ significantly fluctuates for different reionization scenarios. Nevertheless, this latter aspect, if handled properly, can also provide additional constraints during the inference, because the number of ionized regions should be closely connected to the parameters of the ionizing sources \citep{kakiichi2017, giri2018_bsd}. 

Hence, we choose to follow the approach from \citet{vegetti2009} to define our likelihood function. The authors used a strategy that divides the likelihood function into two independent factors: the first one accounting for the likelihood of observing a given number of objects using a Poissonian distribution, and the second to compare the properties of these objects. 

We define a morphological likelihood function, $\mathcal{L}(\text{morphology})$, using two separate factors: a Poissonian factor that compares the number of ionized regions between the model and the observation, and a \textit{distance} factor that compares the two distributions of five dimensional vectors, independently of the number of objects in the distributions. In Appendix~\ref{app:like}, we provide a general expression to define a likelihood function suitable to compare $n$-by-$d$ distributions, where $d$ is the dimension of the statistics used, and $n$ the number of objects that might vary for different sets of parameters. For our case ($d$ = 5), the final likelihood expression is:

\begin{multline}  
\label{eq:likemorph}
\mathcal{L}(\text{morphology}) = \frac{\exp{(-n_{\rm bub}^{\rm mod})}\ \times\  (n_{\rm bub}^{\rm mod})^{n_{\rm bub}^{\rm obs}} }{ n_{\rm bub}^{\rm obs}!} \\ \times  \exp{\left(- \lambda \times \frac{n_{\rm bub}^{\rm mod}}{n_{\rm bub}^{\rm obs}}\sum^{n_{\rm bub}^{\rm obs}}_i \argminA_{j \in\ n_{\rm  mod}} ({\rm d_{\rm maha}} (\vec{v}_{{\rm obs}, i},\vec{v}_{{\rm mod},j})^5)\right)},
\end{multline}

\noindent where $n_{\rm bub}^{\rm mod}$ and $n_{\rm bub}^{\rm obs}$ are the number of ionized bubbles in the observation and in the model, respectively,  $\lambda$ is a regularization parameter, $\rm d_{\rm maha}$ is the Mahalanobis distance \citep{mahalanobis1936}, and $\vec{v}_{{\rm obs}}$ and $\vec{v}_{{\rm mod}}$ are the five dimensional vectors carrying the five morphological attributes for each individual ionized region observed in the observation and model.
A complete description of this likelihood function can be found in Appendix~\ref{app:like}, and we only briefly describe the general idea here. As mentioned above, the factor $\frac{\exp{(-n_{\rm bub}^{\rm mod})}\ \times\  (n_{\rm bub}^{\rm mod})^{n_{\rm bub}^{\rm obs}} }{ n_{\rm bub}^{\rm obs}!}$ assumes that the fluctuation of the number of ionized regions observed follows a Poissonian distribution. The second factor provides a way to compare the distribution of 5D vectors, by finding, for each ionized region in the observation, the ionized structure in the sampled model with similar morphological attributes. This is done by extracting the pair of vectors $\vec{v}_{{\rm mod}}$ and $\vec{v}_{{\rm obs}}$ such that the Mahalanobis distance between them is minimal. The choice of the Mahalanobis distance is motivated by the fact that \textit{classical} distance metrics, such as the Euclidean distance, are not suited for high-dimensional applications \citep{aggarwal2001}. Additionally, the Mahalanobis distance accounts for the covariance of the 5-dimensional distribution of morphological attributes observed (the distributions are normalized to unit variance for each parameters), such that it provides an unbiased metric that is suitable to compare the structures in different distributions of morphological attributes. 

The minimum Mahalanobis distances are computed, elevated to the power five (i.e number of dimensions), and summed for all the ionized regions in the observation. In other words, this can be understood as minimizing the total volume enclosed between the ionized region morphological distribution in the observation and model in a five dimensional space. This approach should favor models whose ionized regions reproduce the observations closest. The factor $\frac{n_{\rm bub}^{\rm mod}}{n_{\rm bub}^{\rm obs}}$ ensures that this second term is normalized to the number of ionized regions in the observations (see details in Appendix~\ref{app:like}). Finally, the parameter $\lambda$ regulates the weight given either to the Poisson factor or to the normalized distance metric that compares the 5D distributions. Optimally, $\lambda$ should be included as an additional parameter during the MCMC inference, such that its best value is determined during the sampling process. However, adding a parameter would further increase the complexity and computational time of the \textsc{21CMMC} run. Hence, in this work, we choose to fix it a priori by performing several test-cases to find an arbitrary optimal value. We find that fixing $\lambda$ to 0.001 provides the best inference results, and we use this value for all results shown in this paper. 

We note that the function defined in Eq~\eqref{eq:likemorph} is not a likelihood function, but rather an approximate penalty function, which is typically suited for regularized maximum likelihood estimation problems. This is because the second factor is not a ``proper'' statistics, contrary to the Poisson factor. Hence, it comes with some caveats, such as the absence of well-defined error terms related to the variance of the 21-cm tomographic statistics with respect to noise or sample variance. Nevertheless, by construction, $\lambda$ can be considered as the inverse variance of the distribution, and therefore, act as a proxy for the error. Additionally, this approach still provide valuable insights to understand whether and how 21-cm tomographic statistics can be included in a Bayesian inference framework, and combined to power spectrum analyses to optimize future observational studies. We will refine this framework in future studies to account for the variance of the 21-cm tomographic statistics. In Section~\ref{sect:results},  we discuss the inference performance of (1) using only the ionized regions number count (the Poissonian factor), (2) using only the ionized regions morphology (the \textit{distance} factor), or (3) using both. Increasing or decreasing $\lambda$ simply shifts the result of (3) towards (2) or (1), respectively.


\subsection{Mock observations}
\label{sect:mock}

Extracting and analyzing the ionized regions from noisy 21-cm images is typically more computationally expensive than deriving the power spectrum. Hence, we choose to focus on relatively small boxes of 128$^3$ cells (250$^3$ Mpc$^3$). We use two sets of mock observations, corresponding to the \textsc{faint sources} and the \textsc{bright sources} models \citep{mesinger2016}, to compare the constraining power of the ionized regions morphological pattern spectra and power spectrum. Additionally, we combine observations at three different redshifts, 10, 9, and 8. The mock power spectra are extracted using the simulated Fourier visibilities and combined with the total theoretical uncertainty computed using the procedure detailed in Section~\ref{sect:likeps}. We then create realistic 21-cm mock images including the SKA1-Low PSF, a random noise realization, and the additional 3D smoothing. 

The models are sampled using a different set of initial conditions (i.e different realization of the density field) on the same grid size as the mock observations. Studies using the power spectrum typically use a larger box size to create the mock observations, while the models are produced on a smaller grid (but keeping the same cell resolution) \citep[e.g.][]{greig2015,greig2017}. Nevertheless, 21-cm tomographic statistics are not independent of the box size used (e.g. the number of ionized structures will increase for larger boxes), thus, we keep the same grid for both the models and the mock observations. We note that our experiment is likely more affected by sample variance due to this relatively small box size, and we further discuss this point in Section~\ref{sect:sample}. 

As detailed in Section~\ref{sect:noisepsf}, both the noise and the PSF impact the number and morphological attributes of the ionized regions. Consequently, these instrumental effects must also be included when sampling the parameter space during the inference process, to consistently compare models with observations. During the MCMC sampling process, the ionized regions number-count and morphological spectra of each model sampled are extracted on the fly by (1) using the \textit{Triangle Thresholding} approach to segment the ionized regions from the mock noisy, low-resolution 21-cm tomographic image cubes, and (2) applying \textsc{DISCCOFAN} on the resulting segmented volumes (similar to Section~\ref{sect:noisepsf}). We note that the location of the ionized regions is always extracted directly from the mock 21-cm observations, such that, in principle, the same approach can be applied on real observational data sets. To accelerate the convergence of the MCMC analysis, we keep the noise realization fixed for each model sampled (but chosen different than for the mock observations). We performed several test-cases to ensure that the choice of a particular noise realization has little impact on the final inference results.  The stopping criterion in \textsc{21CMMC} is defined relative to a certain number of sample iterations, rather than to a convergence criterion. Therefore, we test for the convergence of the Markov chains using two tests. The first one is the Gelman-Rubin diagnostic \citep{gelman1992}, which computes the sample mean and variance from multiple chains, and check whether they are similar enough to indicate approximate convergence. Additionally, we also use the Geweke diagnostic \cite{geweke1991}, which compares the similarity of the mean and variance of segments from the beginning and end of a single chain. We note that these two tests \textit{estimate} the convergence of the chains but are no proof of actual convergence. They are usually used to prove a failure to converge, rather than guaranteeing that the chains converged to a global minimum.

Overall, running \textsc{21CMMC} using the 21-cm tomographic statistics, 3000 iterations and 50 walkers takes around 8 days on a machine with 80 CPUs. Using only the power spectrum with the same setup takes approximately a factor two less time. Assessing the performance of this approach on larger observations will be crucial to understand the full potential of 21-cm tomographic statistics. To do so, using emulators of the tomographic 21-cm image cubes, such as in \citet{chardin2019} and \citet{list2020}, might prove useful to extend this work.  

\section{Results}
\label{sect:results}
This section presents the inference results obtained with \textsc{21CMMC}. Section~\ref{sect:resmorph} details the recovered parameter intervals using the 21-cm tomographic statistics for both the \textsc{faint sources} and \textsc{bright sources} mock observations. Section~\ref{sect:resps} compares these results to the constraints inferred using the power spectrum of the 21-cm fluctuations. Finally, Section~\ref{sect:sample} explores the impact of using different initial conditions, and Section~\ref{sect:gauss} investigates the consequences of including simulated foreground residuals, as one of the dominant systematic errors, on these results.

\subsection{Inference using the ionized regions number-count and morphology}
\label{sect:resmorph}

\begin{figure*}  
\includegraphics[width = 0.8\textwidth]{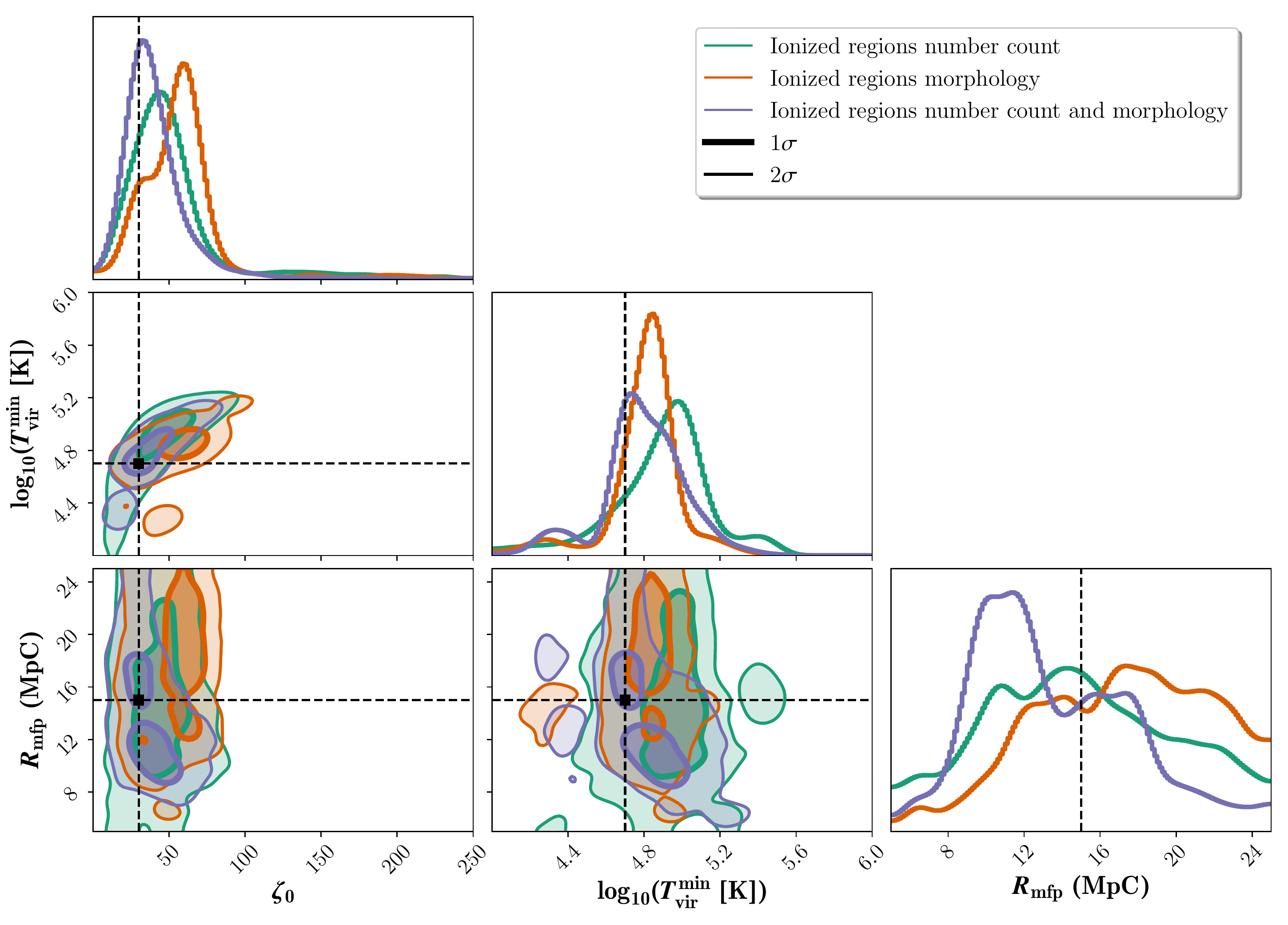}
 \caption{The 1D marginalized PDFs (diagonal panels) and 2D likelihood contours (lower panels) of the three astrophysical parameters  $\zeta$, log$_{10}(T^{\rm min}_{\rm vir})$ and $R_{\rm mfp}$ for the \textsc{faint sources} mock observations assuming 1000h observations with SKA at redshifts 10, 9  and 8. The dashed lines show the fiducial parameters ($\zeta$ = 30, log$_{10}(T^{\rm min}_{\rm vir})$ = 4.7 and $R_{\rm mfp}$ = 15 Mpc). We compare three MCMC cases using: the number of ionized regions (green); the morphology of the ionized regions (orange); and combining both (purple). }
 \label{fig:30_hii_morph}
\end{figure*}

\begin{figure*}  
\includegraphics[width = 0.8\textwidth]{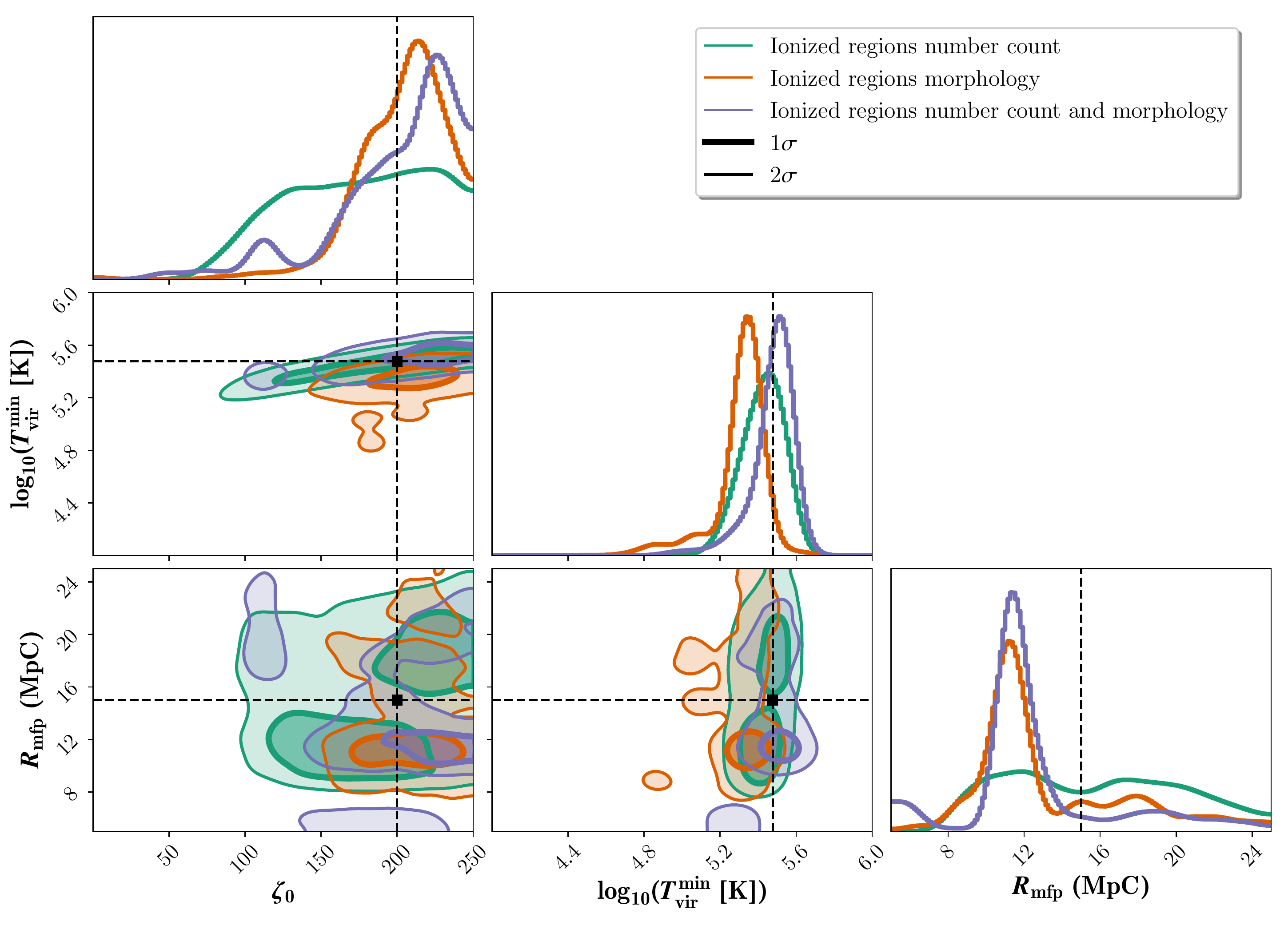}
 \caption{Same as Figure~\ref{fig:30_hii_morph} for the \textsc{bright sources} mock observations ($\zeta$ = 200, log$_{10}(T^{\rm min}_{\rm vir})$ = 5.48 and $R_{\rm mfp}$ = 15 Mpc).}
 \label{fig:200_hii_morph}
\end{figure*}

\begin{table*} 
	\centering

	\label{tab:mcmc_morph}
	\begin{tabular}{ccccc} 
		\hline
		Case / Method & $\zeta$ & log$_{10}(T^{\rm min}_{\rm vir} [K])$ & $R_{\rm mfp}$ (Mpc)\\
		\hline		\hline
		\textsc{faint sources} & 30.00 & 4.70 & 15.00 \\
		\hline
		Ionized regions number count& $45.19^{+19.57}_{-17.23}$ & $4.93^{+0.15}_{-0.27}$ & $14.66^{+5.85}_{-4.71}$\\
		Ionized regions morphology& $57.52^{+12.37}_{-26.52}$ & $4.83^{+0.08}_{-0.13}$ & $17.32^{+4.68}_{-5.13}$\\
		Ionized regions number count and morphology & $35.02^{+18.36}_{-7.39}$ & $4.78^{+0.19}_{-0.12}$ & $12.54^{+5.26}_{-2.84}$\\
		Power spectrum & $31.29^{+1.24}_{-1.17}$ & $4.74^{+0.03}_{-0.03}$ & $22.11^{+1.76}_{-2.31}$\\
		\hline
			\textsc{bright sources}& 200.00 & 5.48 & 15.00 \\
		\hline
		Ionized regions number count & 176.85$^{+48.81}_{-55.26}$ & $5.44^{+0.08}_{-0.11}$ & $15.57^{+5.17}_{-5.01}$\\
		Ionized regions morphology & $208.22^{+21.54}_{-30.85}$ & $5.34^{+0.04}_{-0.06}$ & $11.53^{+6.40}_{-0.89}$\\
		Ionized regions number count and morphology & $211.38^{+23.34}_{-47.12}$ & $5.51^{+0.02}_{-0.11}$ & $11.55^{+5.07}_{-0.60}$\\
		Power spectrum & $208.68^{+20.27}_{-20.55}$ & $5.50^{+0.03}_{-0.03}$ & $17.43^{+1.62}_{-2.14}$\\
	\end{tabular}
		\caption{The median values, and the associated 16$^{\rm th}$ and 84$^{\rm th}$ percentile errors of the recovered astrophysical parameters, $\zeta$,  log$_{10}(T^{\rm min}_{\rm vir})$, and $R_{\rm mfp}$ for the two different reionization scenarios. We assumed 1000h of observations with SKA1-Low for three different epochs at $z$ = 10, 9 and 8. We compare the inference results using: only ionized regions number count; only the ionized regions morphology; both the ionized regions number-count and morphology; and the power spectrum.}
\end{table*}

We first assess the performance of the 21-cm tomographic inference approach to recover the parameters of the sources of cosmic reionization. In Figure~\ref{fig:30_hii_morph} and Figure~\ref{fig:200_hii_morph}, we compare the outputs of \textsc{21CMMC} for the \textsc{faint sources} and \textsc{bright sources} models, respectively, considering three different cases based on the likelihood function defined in Eq.~\eqref{eq:likemorph}: (1) using the number of ionized regions observed (i.e the Poissonian factor); (2) using the morphology of the ionized regions (i.e the distance factor); and (3) combining (1) and (2). The results are shown as corner plots\footnote{Corner plots from Figures~\ref{fig:30_hii_morph}, \ref{fig:30_hii_ps}, \ref{fig:200_hii_morph}, \ref{fig:200_hii_ps} and \ref{fig:30_hii_gauss} have been made using the python package \textsc{corner} \citep[][https://github.com/dfm/corner.py/blob/main/docs/index.rst]{corner}}, with the diagonal panels providing the normalized marginalized 1D PDFs of the three parameters, and the lower panels representing the 2D likelihood contours, at the 1 and 2 $\sigma$ level (thick and thin lines, respectively). Additionally, Table~\ref{tab:mcmc_morph} shows the recovered median values and the associated 16$^{\rm th}$ and 84$^{\rm th}$ percentile errors, assuming that our framework, and in particular the factor 1/$\lambda$, decently traces the variance for the attribute values (see discussion in Section~\ref{sect:likemorph}). In the following sub-sections, we detail the results of each inference case.

\subsubsection{Comparing only the number of ionized regions}
\label{sect:hiinumber}
Using the number of ionized regions information already provides remarkably good constraints for the recovered ionizing efficiency and the minimal virial temperature for both reionization scenarios. We note that the median of the posterior distributions are slightly offset compared to the fiducial parameters (around +15 for $\zeta$ and +0.2 for log$_{10}(T^{\rm min}_{\rm vir})$ for the \textsc{faint sources}, and around $-25$ for $\zeta$ for the \textsc{bright sources} model), but these fairly accurate intervals already support that this information is useful to infer information about the ionizing sources. This is not unexpected since both the values of $\zeta$ and $T^{\rm min}_{\rm vir}$ drive the number of ionized regions. As discussed in Sect.~\ref{sect:model}, models with lower $T^{\rm min}_{\rm vir}$  have fewer ionized regions because the ionizing sources can only exist within the densest halos. Similarly, a larger $\zeta$ will boost the number of ionizing photons produced by the sources, accelerating the pace at which ionized bubbles grow, and thus increasing the merging rate during the early stages. In both cases, the observed number of individual ionized regions should decrease. Consequently, this information can already rule out a large number of models that have diverging $n^{\rm mod}_{\rm bub}$ and $n^{\rm obs}_{\rm bub}$.  The 2D likelihood contours in Figure~\ref{fig:30_hii_morph} show the degeneracy between the ionizing efficiency and the minimum mass of the star-forming halos. In the \textsc{faint sources} model, the recovered number of ionized regions does not significantly vary for models with larger $\zeta$ and log$_{10}(T^{\rm min}_{\rm vir})$, while for the \textsc{bright sources} model, large fluctuations of the ionizing efficiency of the sources still produce a similar number of ionized regions, but small variations  of $T^{\rm min}_{\rm vir}$ have a significant impact on $n^{\rm obs}_{\rm bub}$. The degeneracy between $\zeta$ and $T^{\rm min}_{\rm vir}$ is expected because they both impact the number of ionized regions in each model, which depends on the criterion used in Equation~\eqref{eq:ionizcrit} ($\zeta\times f_{\rm coll}$) that sets the ionized fraction in each cell. In Section~\ref{sect:indiv}, we further demonstrate that the 2D likelihood contours of $\zeta$ and $T^{\rm min}_{\rm vir}$ almost exactly follow the isocontours of the average $\zeta\times f_{\rm coll}$ for a given reionization scenario.

We note that the number of ionized regions does not provide enough information to accurately infer the mean free path of the ionizing photons for both the  \textsc{faint sources} and \textsc{bright sources} models. While the $R_{\rm mfp}$ fiducial value is within the recovered interval, the large fractional errors show that this parameter is not tightly constrained. In theory, $R_{\rm mfp}$ regulates the maximum scale to which an ionized region can grow around the ionizing sources, and becomes important only when the ionized regions grow larger than this value. \citet{greig2017} emphasized that its impact on the power spectrum is less significant when its value is larger than $\sim$15 Mpc because the merging of the ionized bubbles becomes the most dominant growth mechanism. Our result suggests that only models with values lower than 8 Mpc seem to be robustly ruled out. Overall, the impact of $R_{\rm mfp}$ fluctuations is likely too complex to recover for our experimental setup given the limited box size and low-resolution aspect of the data sets (see discussion in Section~\ref{sect:sample}).

We note that changing the box size or its resolution will impact the number of ionized regions observed and could affect these results. Typically, when using larger box sizes, we expect to rule out more efficiently astrophysical models that do not closely reproduce the number of ionized regions in the mock observations. This is because increasing $n_{\rm bub}$ typically causes larger variations in the Poisson likelihood factor, which significantly increase the differences in the log-likelihood values. However,  to accurately quantify these effects, we need to test our inference framework on larger boxes. This is currently computationally not feasible on machines available to us, but will be investigated in future works.

\subsubsection{Comparing only the ionized regions morphology}

Comparing the morphological properties of the ionized regions also provides valuable information to constrain $\zeta$ and $T^{\rm min}_{\rm vir}$ for both reionization models. For the \textsc{faint sources} model, the recovered ionizing efficiency of the sources is slightly offset (around $+27$), while the minimum virial temperature is more robustly inferred. On the other hand, $\zeta$ is better constrained for the \textsc{bright sources} case, but $T^{\rm min}_{\rm vir}$ interval is shifted by around $-0.14$ (logarithmic). The recovered values are still at $\pm 3\sigma$, however, given the particular definition of the likelihood function implemented, the robustness of these uncertainties should be taken with caution. Overall, these results suggest that the morphological properties of the ionized structures also provide independent information to recover the properties of the underlying reionization sources. The accurate inference of $T^{\rm min}_{\rm vir}$ is slightly surprising since the virial temperature regulates the number of ionizing sources formed by selecting the mass threshold of the halos that can form such sources. Therefore, it is unclear how this parameter is connected to the morphological properties of the sources. Nevertheless, the \textit{distance} factor in Equation~\eqref{eq:likemorph} ensures that the models with the best likelihood values have similar morphological pattern spectra than in the observation. Hence, models with different $T^{\rm min}_{\rm vir}$  values might be efficiently ruled out because they do not produce ionized regions with the same morphological spectra. In Section~\ref{sect:indiv}, we actually show that the ionized regions morphology is indirectly pre-determined by the average ionizing budget per baryon, which depends on both the values $\zeta$ and $T^{\rm min}_{\rm vir}$. 

We note in Figure~\ref{fig:30_hii_morph} that the morphology of the ionized regions does not help to place a robust constraint on $R_{\rm mfp}$, but can only be used to rule out models with $R_{\rm mfp}$ < 8 Mpc. In theory, fluctuations in $R_{\rm mfp}$ impact the volume of the ionized structures, such that using the size distribution of the  ionized regions should provide information to infer this IGM property. Nevertheless, as mentioned already in Section~\ref{sect:hiinumber}, these differences might be too complex to accurately measure given our experimental setup. 

\subsubsection{Combining the number-count and morphology of the ionized regions}

Finally, combining the number count and the morphological attributes of the ionized regions provides the best inference results for the joint set of values of $\zeta$ and $T^{\rm min}_{\rm vir}$ for both the \textsc{faint sources} and \textsc{bright sources} models. The posterior distributions peak closer to the fiducial values, and have smaller 16$^{\rm th}$ and 84$^{\rm th}$ percentile errors. On the other hand, it does not improve the inferred interval of $R_{\rm mfp}$, where the 2D likelihood contours are typically extended over the whole $R_{\rm mfp}$ prior range. In Section~\ref{sect:sample}, we highlight that the recovered interval is strongly affected by the set of initial conditions. Nevertheless, we expect this outcome to be different for experiments where both the observation and the models are sampled on a larger grid because fluctuations in the $R_{\rm mfp}$ values should be more significantly observable based on the morphological properties or the number of the ionized structures. 

Overall, Figures~\ref{fig:30_hii_morph} and \ref{fig:200_hii_morph} highlight that the combination of the number and morphological attributes of the ionized structures extracted from 21-cm images provides an alternative approach to the power spectrum to disentangle different EoR scenarios, and constrain properties of the ionizing sources. This first test-case paves the way for more complex, and deeper explorations of these alternative approaches using Bayesian inference frameworks.

\subsection{Comparison with the power spectrum}
\label{sect:resps}

\begin{figure*}  
\includegraphics[width = 0.8\hsize]{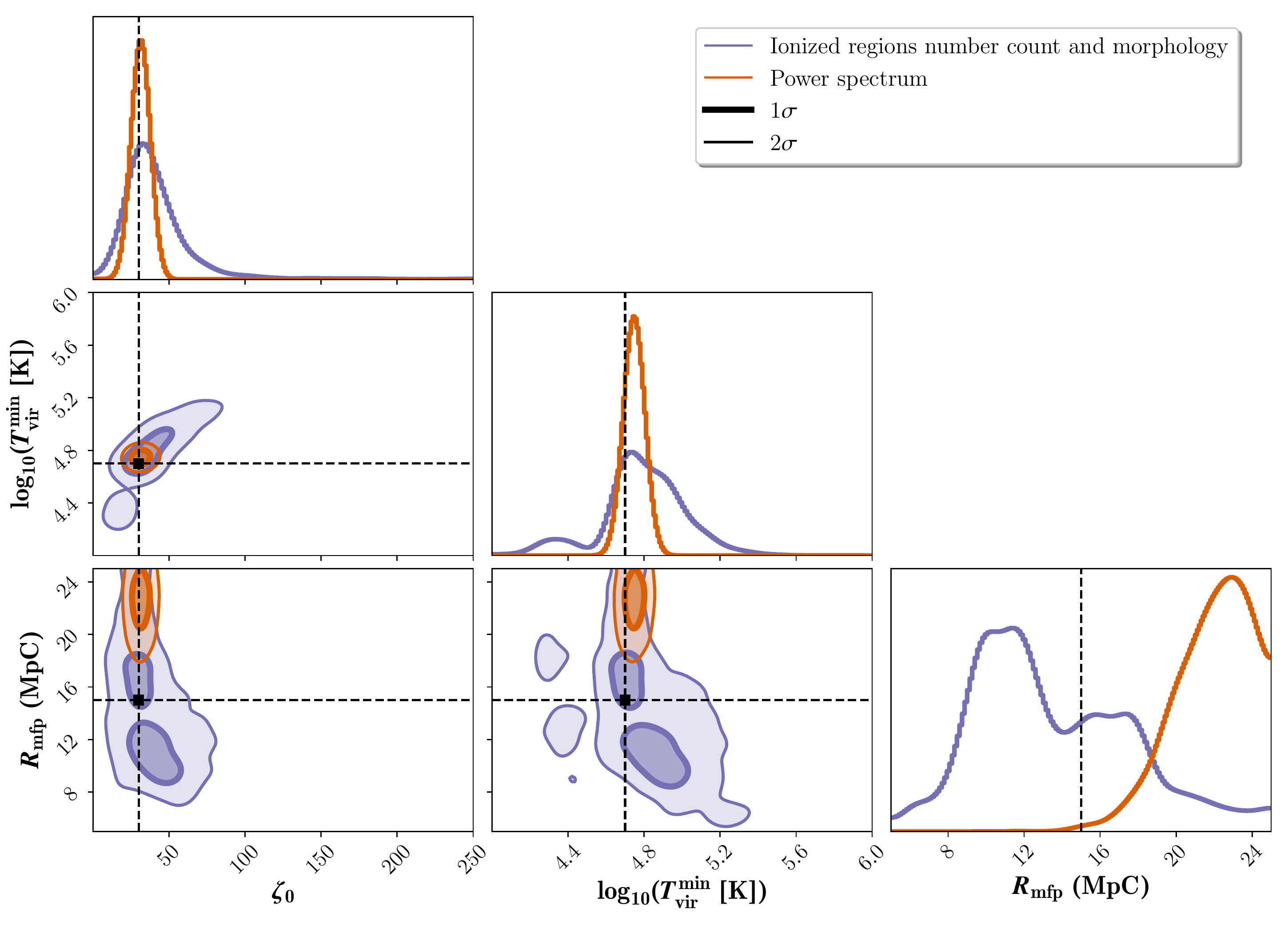}
 \caption{The 1D marginalized PDFs (diagonal panels) and 2D likelihood contours (lower panels) of the three astrophysical parameters  $\zeta$, log$_{10}(T^{\rm min}_{\rm vir})$ and $R_{\rm mfp}$ for the \textsc{faint sources} model assuming 1000h observations with SKA at redshifts, 10, 9  and 8. The dashed lines show the fiducial parameters ($\zeta$ = 30, log$_{10}(T^{\rm min}_{\rm vir})$ = 4.7 and $R_{\rm mfp}$ = 15 Mpc). We compare the outcomes of 21CMMC using the power spectrum (orange) or the number and morphology of the ionized regions (purple).}
 \label{fig:30_hii_ps}
\end{figure*}

\begin{figure*}  
\includegraphics[width = 0.8\hsize]{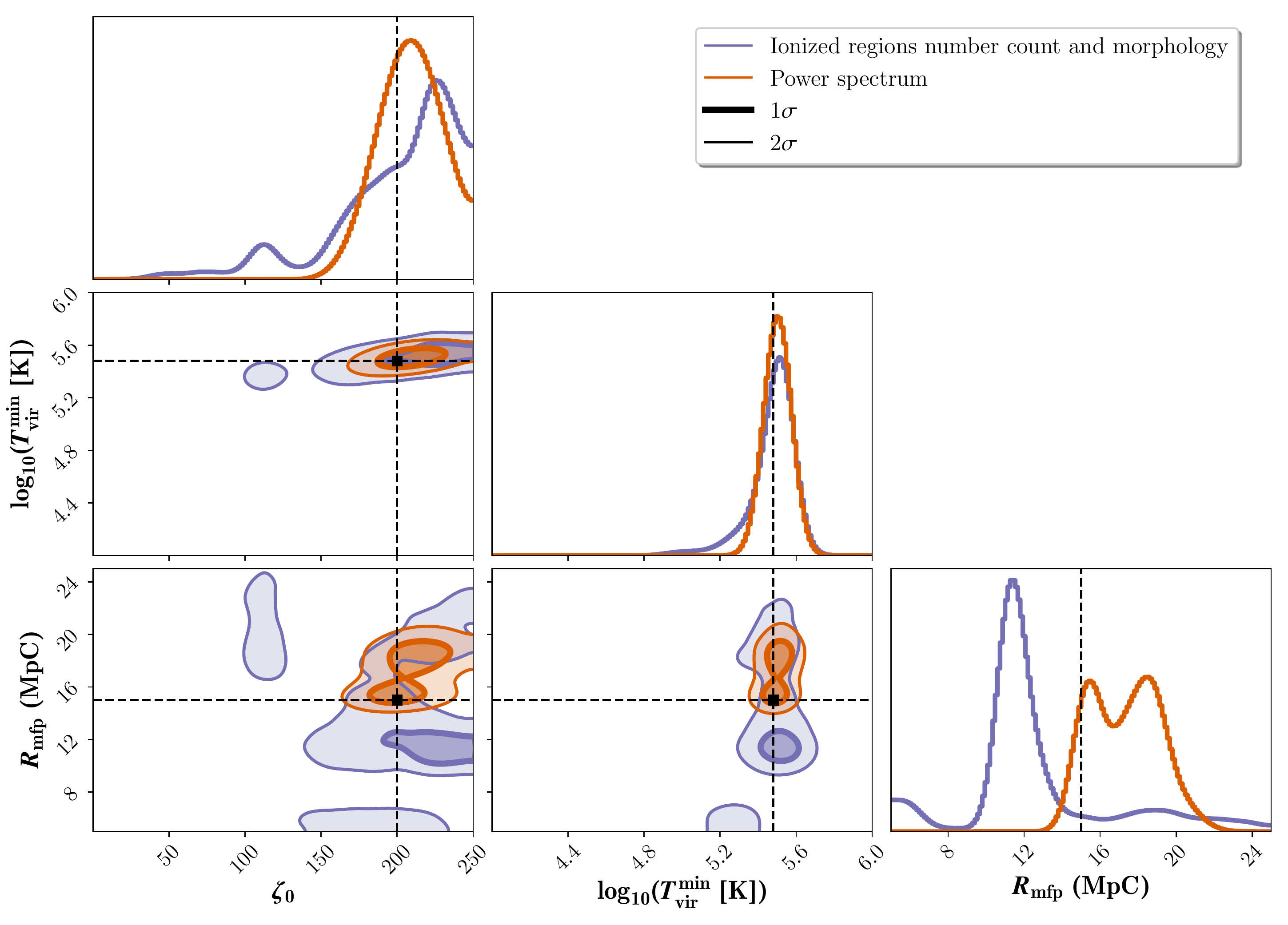}
 \caption{Same as Figure~\ref{fig:30_hii_ps} for the \textsc{bright sources} model (fiducial parameters $\zeta$ = 200, log$_{10}(T^{\rm min}_{\rm vir})$ = 5.48 and $R_{\rm mfp}$ = 15 Mpc).}
 \label{fig:200_hii_ps}
\end{figure*}

Several studies \citep[e.g][]{greig2015, greig2017, greig2018, park2019, Binnie2019} showed that a Bayesian statistical framework using the power spectrum would already provide tight constraints on the EoR for different reionization parameters or models. However, the power spectrum is only a limited statistic, because it does not encode the phase information from the Fourier modes, nor the information about non-Gaussianities or non-ergodic effects. In theory, 21-cm tomographic images give a more complete description of the evolution of the ionizing fronts and probe additional astrophysical properties through the non-gaussianities of the 21-cm signal. In this section, we simply explore how the constraints inferred using the 21-cm tomographic statistics compare to the ones set by the power spectrum of the 21-cm observations.
 
In Figures~\ref{fig:30_hii_ps} and \ref{fig:200_hii_ps}, we compare the inferred 1D PDFs and 2D likelihood contours when using the power spectrum or the ionized regions number count and morphological spectra for the \textsc{faint sources} and \textsc{bright sources} models, respectively. Additionally, the corresponding medians and 16$^{\rm th}$ and 84$^{\rm th}$ percentile errors can be found in Table~\ref{tab:mcmc_morph}. Overall, the power spectrum provides tighter constraints on $\zeta$ and $T^{\rm min}_{\rm vir}$ compared to the 21-cm tomographic statistics. These results are not unexpected since, as shown in Appendix~\ref{app:ps},  the theoretical uncertainties on the mock power spectrum used in this work are relatively small at most of the $k$ modes sampled, which strongly rules out all models that do not closely reproduce its shape. Additionally, the power spectrum is extracted directly from the simulated brightness temperature fluctuations in the Fourier space, while the ionized morphological attributes are extracted from noisy, lower resolutions, image cubes. Hence, these tomographic statistics are more sensitive to the quality of the 21-cm images, which depends on the additional processing steps, such as the choice of the smoothing kernel, the uv weighting scheme, or the robustness of the ionized regions extraction method.

\begin{figure}  
\includegraphics[width = \hsize]{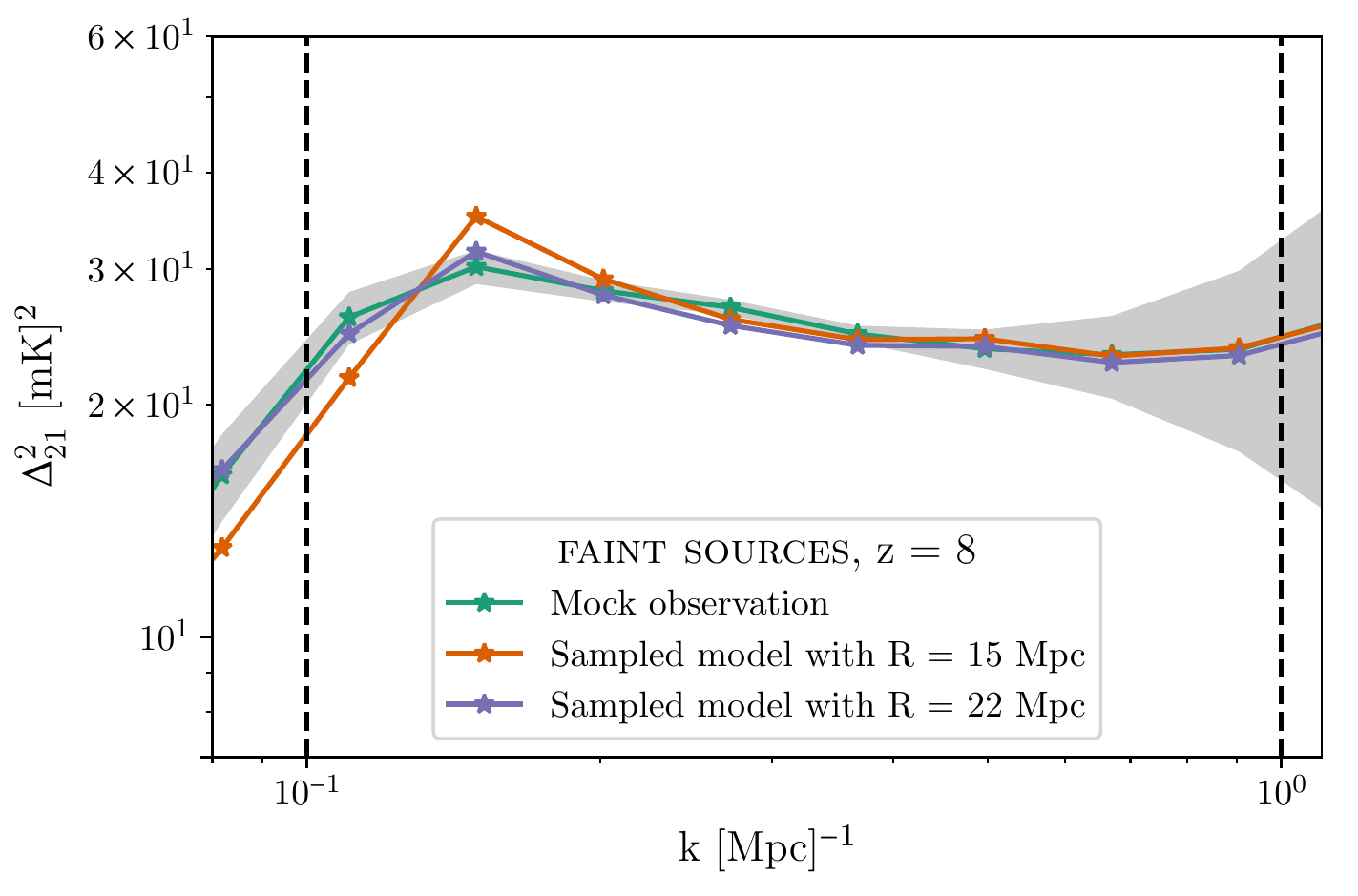}
 \caption{The power spectrum of the mock observation for the \textsc{faint sources} model (green), the sampled model with same parameters as the mock observation (orange), and the sampled model with $R_{\rm mfp}$ = 22 Mpc  (purple). The differences at low k (large scales) for the same set of parameters but different initial conditions suggest that our experiment is significantly impacted by sample variance.}
 \label{fig:wrong_ps}
\end{figure}

We note that using the power spectrum only provides a fairly accurate interval for $R_{\rm mfp}$ for the \textsc{bright sources} case, but not for the \textsc{faint sources} case where the median recovered value is at more than 3$\sigma$ than the fiducial value. This is likely because our experiment is significantly impacted by sample variance due to the small box sizes used. In Figure~\ref{fig:wrong_ps}, we show the spherically averaged power spectrum extracted at redshift $z$ = 8 from the mock observation of the \textsc{faint sources} set of parameters, the power spectrum from the sampled model with the same fiducial set of parameters (but different initial conditions), and the power spectrum from the model with the fiducial $\zeta$ and $T^{\rm min}_{\rm vir}$ but $R_{\rm mfp}$ = 22. It highlights that the power spectrum of the model with the same set of fiducial parameters significantly differs at the low $k$ modes (which probe the large scales) compared to the mock observation, such that models with a larger $R_{\rm mfp}$ (22 Mpc) better reproduce the power spectrum of the mock observation.  Typically, the $R_{\rm mfp}$ fluctuations dominantly affect the large scales \citep[as shown in Figure 2 of][]{greig2015}, which can be strongly affected by statistical variance in boxes with limited size \citep{Kaur2020}. This caveat is further investigated in Section~\ref{sect:sample}, where we compare the outputs of 10 MCMC runs with different sets of initial conditions.

Overall, these results suggest that using the power spectrum provides tighter constraints on the ionizing efficiency and the minimum mass of the star-forming halos compared to statistics derived on noisy, lower resolution 21-cm images. While this might question the use of these statistics to infer the astrophysics of reionization, we show in Section~\ref{sect:gauss} that 21-cm tomographic statistics are more robust to the presence of residual artifacts in the observation. Additionally, we note that we compared, but never combined both statistics within the same inference analysis. This is because these approaches are not entirely independent from each other since they both encode for example the typical scale of the ionized structures. We discuss in Section~\ref{disc:morphuse} how alternative 21-cm tomographic or higher-order statistics could be further combined to a power-spectrum analysis to improve the inference results.

\subsection{Impact of different initial conditions}
\label{sect:sample}
\begin{figure*}  
\includegraphics[width = 0.8\textwidth]{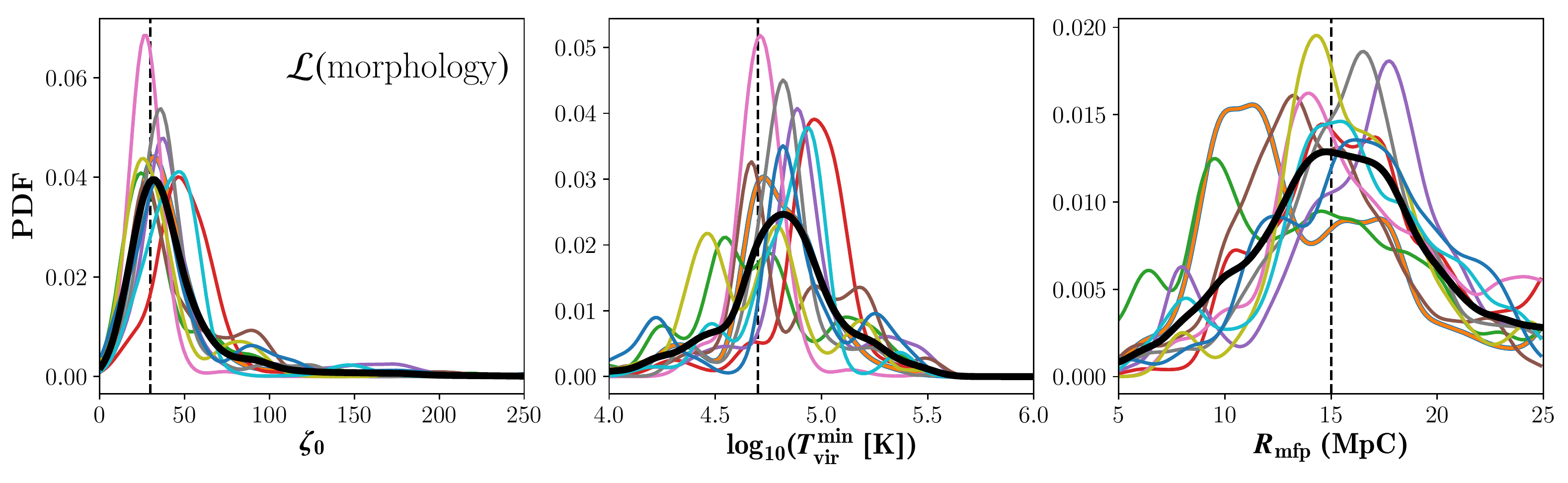}
\includegraphics[width = 0.8\textwidth]{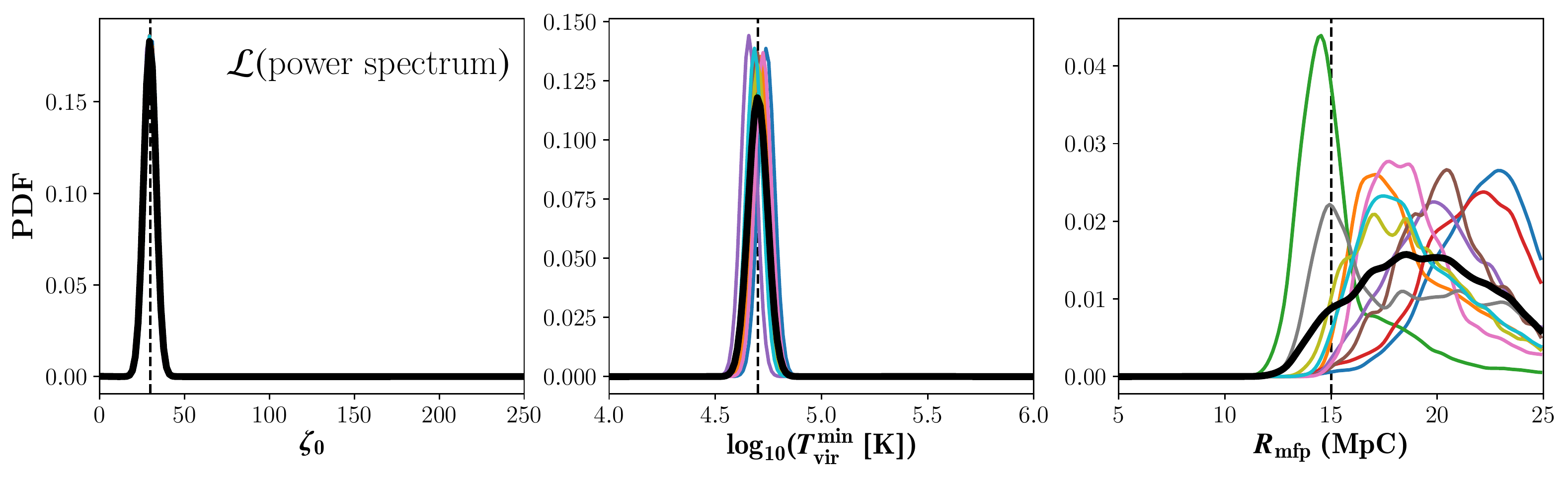}
 \caption{Top: the 1D marginalized PDFs for the three astrophysical parameters  $\zeta$, log$_{10}(T^{\rm min}_{\rm vir})$ and $R_{\rm mfp}$ using the  likelihood function based on the number count and morphology of the ionized regions and the \textsc{faint sources} mock observations. Bottom: same but using the likelihood function based on the power spectrum. We compare ten MCMC runs corresponding to ten different sets of initial conditions (thin color lines). The black thick line corresponds to the average of the ten runs.}
 \label{fig:30_hii_morphcum}
\end{figure*}


In \textsc{21cmFAST}, the set of initial conditions determines the underlying density field, which is then used to compute the brightness temperature fluctuations at different redshifts. In this work, we fixed the initial conditions of the sampled models a priori, but choosing it different than the ones used to create the mock observations. A more robust inference framework would consist in varying the initial conditions as we sample the models during the inference, nevertheless, this is computationally costly as it requires to re-compute the evolved density field for each different set of parameters. As already mentioned in the previous section, the choice of the initial conditions can lead to significant differences when the box size is limited. Recently, \citet{Kaur2020} quantified the error using a mock power spectrum extracted from a box of length 1.1 Gpc and showed that boxes larger or equal than 250 Mpc are needed to achieve convergence within 1$\sigma$ level of the noise. For our experiment, both the mock observation and the models are extracted on boxes with this limited physical size, such that we might be more sensitive to this statistical variance. Hence, in this section, we aim to quantify the impact of sample variance on the previous results by running ten different \textsc{21CMMC} runs using ten different sets of initial conditions for the sampled models. Because running these tests is computationally expensive (around seven days per run), we perform this test only for the \textsc{faint sources} model, and use a lower number of iterations for the MCMC runs.

The top and bottom panels of Figure~\ref{fig:30_hii_morphcum} show the results of our experiment using the ionized region number count and morphological spectra, and using the power spectrum, respectively. For clarity, we only show the 1D posterior PDFs, since the 2D likelihood contours do not provide useful information. The black line in each figure represents the average of the ten 1D PDFs. We note that these averaged PDFs are not physically motivated, as they do not represent the combination of ten independent results (for example by observing ten different fields), but simply summarize the typical impact of varying initial conditions. Additionally, we report in Appendix~\ref{app:tablemcmc} the median and 16$^{\rm th}$ and 84$^{\rm th}$ percentiles for all ten cases. 

Overall, the top panels in Figure~\ref{fig:30_hii_morphcum} support the findings from Section~\ref{sect:resmorph}. The ionized morphological and number count information provide fairly accurate constraints for $\zeta$ and $T^{\rm min}_{\rm vir}$, such that the interval of median values recovered from the posterior distributions in each case is between 28.28 and 49.38 for the former, and between $ 5.0\times10^4$ and $9.0\times10^4$ K for the latter. Interestingly, these values are all within 1.5$\sigma$ of the results presented in Section~\ref{sect:resmorph}, suggesting that systematic errors due to sample variance are consistently contained within the width of the recovered parameter interval from independent runs. We also note the degeneracy between the ionizing efficiency and the minimum virial temperature, already highlighted in Section~\ref{sect:resmorph}, such that a larger recovered $\zeta$ is typically paired with a larger $T^{\rm min}_{\rm vir}$. The origin of this degeneracy is further discussed in Section~\ref{sect:indiv}.

Additionally, the top panels in Figure~\ref{fig:30_hii_morphcum} emphasizes that the width and the peak of the posterior distribution of $R_{\rm mfp}$ significantly vary depending on the set of initial conditions when using the likelihood function based on the number and morphological spectra of the ionized regions. On average, models with very low or very large $R_{\rm mfp}$ seem slightly more disfavored, but the large percentile errors prevent us from setting a robust constraint to this IGM property in all individual cases. 

When considering the power spectrum, Figure~\ref{fig:30_hii_morphcum} emphasizes that varying the initial conditions has little impact on the recovered ionizing efficiency and minimum virial temperature. The interval of the lower and higher recovered median values is within 29 and 31 for $\zeta$ and $4.7\times10^4$ and $5.5\times10^4$ for $T^{\rm min}_{\rm vir}$, and all results are within 1.5$\sigma$ from the interval obtained in Section~\ref{sect:resmorph}. This suggests that constraints on both parameters are robust against the effect of statistical variance. Nevertheless, this is not the case for $R_{\rm mfp}$. The ten posterior distributions have significant differences depending on the set of initial conditions used, such that the average interval of accepted values is within $\sim$ 14 and 25 Mpc. This is consistent with \citet{greig2015} who show that this parameter has a stronger impact on the growth of the ionized regions for small values, while it leads to little difference when larger than $\approx 15$ Mpc. The results of the ten runs suggest that this parameter is adjusted to match the realization of the large scale structure from the mock observation.

As discussed before, this caveat can be solved by using a larger grid for both the mock observations and the models sampled, to reduce the statistical variance impacting the large scales. Additionally, it shows that 21-cm tomographic statistics, albeit extracted from noisy and lower resolution images, are still fairly robust against sample variance effects. However, the constraints inferred using the power spectrum seems more precise than using the number and morphological properties of the ionized regions. Nevertheless, we show in the next section that the power spectrum is likely more sensitive to the presence of foreground residuals compared to statistics extracted from 21-cm images, and thus could be less accurate.  

\subsection{Impact of foreground residuals}
\label{sect:gauss}

\begin{figure*}  
\includegraphics[width = \hsize]{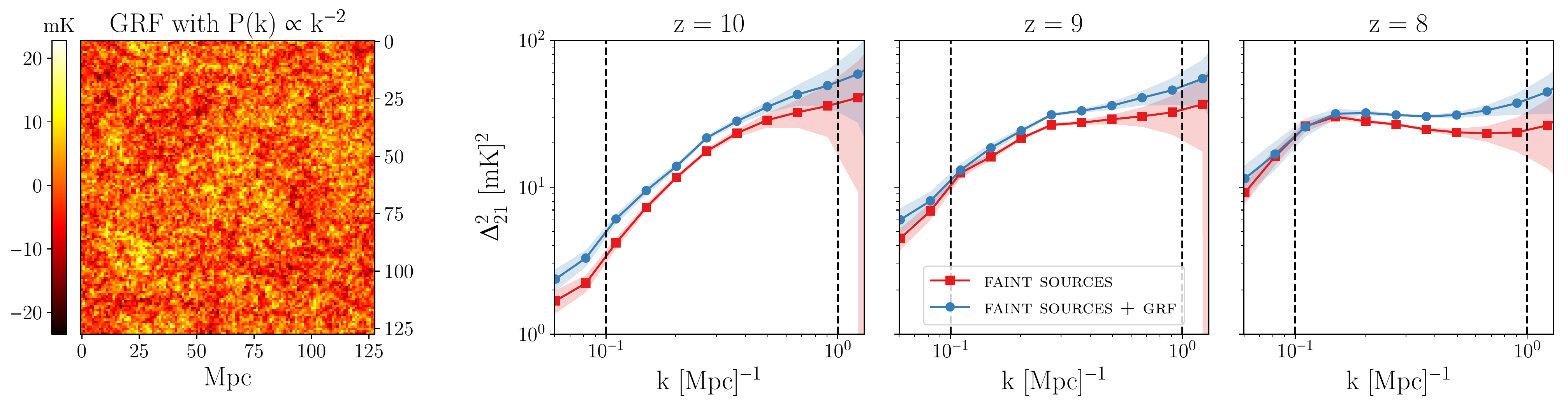}
 \caption{From left to right: A 2D slice of the Gaussian Random Field (GRF), mimicking foreground residuals, added to the observation at redshift 10; the corresponding power spectrum of the original mock model and of the observation + GRF at redshift 10, 9, and 8.}
 \label{fig:img_gauss}
\end{figure*}

\begin{table}
	\centering
	\label{tab:mcmc_gauss}
	\begin{tabular}{ccccc} 
		\hline
		Method & $\zeta$ & log$_{10}(T^{\rm min}_{\rm vir} [K])$ & $R_{\rm mfp}$ (Mpc)\\
		\hline		\hline
		Tomography & $46.88^{+7.15}_{-9.14}$ & $4.90^{+0.06}_{-0.12}$ & $13.89^{+3.98}_{-3.82}$\\
		Power spectrum & $71.16^{+4.57}_{-4.21}$ & $5.17^{+0.03}_{-0.03}$ & $23.56^{+0.91}_{-1.65}$\\
		\hline
	\end{tabular}
		\caption{The median values, and the associated 16$^{\rm th}$ and 84$^{\rm th}$ percentile errors for the three astrophysical parameters, $\zeta$,  log$_{10}(T^{\rm min}_{\rm vir})$, and $R_{\rm mfp}$ for the modified \textsc{faint sources} model ($\zeta$ = 30, log$_{10}(T^{\rm min}_{\rm vir})$ = 4.7 and $R_{\rm mfp}$ = 15 Mpc) perturbed with a 3D Gaussian Random Field (see Figure~\ref{fig:img_gauss}). }
\end{table}

In Sections~\ref{sect:resps} and \ref{sect:sample}, we showed that 21-cm tomographic statistics provide an alternative approach to the power spectrum to disentangle different reionization scenarios. However, using the power spectrum sets tighter constraints on the parameters of the reionization sources. As we already discussed, this is likely because the 21-cm tomographic statistics only account for the properties of the ionized regions, and are extracted on lower resolution data sets that require additional processing steps. However, 21-cm signal images provide a direct measure of the ionized structures and their properties, while the power spectrum is an averaged statistics. Consequently, the presence of diffuse structures locally have a limited impact on the shape of the ionized regions, while it will affect the power spectrum on large scales. In this section, we investigate this effect using simulations of three dimensional Gaussian random fields (GRF) having a power spectrum consistent with a power-law k$^{\alpha}$ with $\alpha = -2.0$ \citep[consistent with observations of large scale diffuse Galactic emission using the Amsterdam-ASTRON Radio Transient Facility and Analysis Centre (AARTFAAC);][]{ghelot2019} and a brightness temperate variance of 28.9 mK.  We add these GRF to the three mock observations of the \textsc{faint sources} model at redshifts 10, 9 and 8. These GRFs can be seen as diffuse structures that have not been correctly removed during the foreground subtraction.  Figure~\ref{fig:img_gauss} shows a 2D slice of one of the 3D GRF realization, and the mock power spectrum extracted from the modified observations. The \textit{contaminated} power spectra are shifted towards larger values, and this effect is stronger for larger $k$ bins. This is expected since adding a GRF will enhance the variance if the field is uncorrelated with the 21-cm field. We note that our simulation of the GRF assumes that it fluctuates similarly in the frequency and spatial domain. The latter is not very realistic since these diffuse residuals are typically synchrotron spectra, which are frequency coherent. Nevertheless, it still provides a decent test-case to investigate the impact of foreground contamination.

\begin{figure*}
\includegraphics[width = 0.8\hsize]{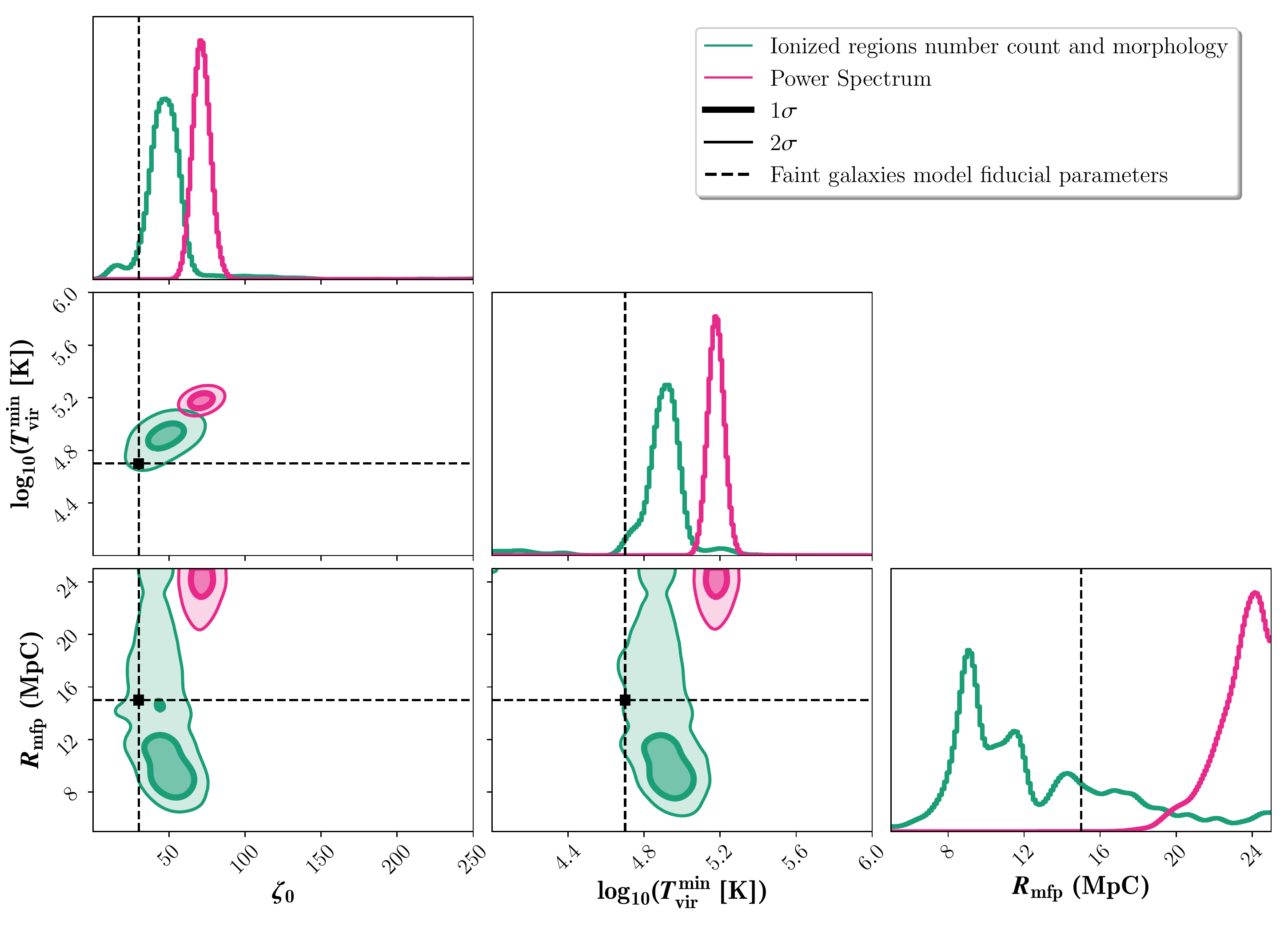}
 \caption{The 1D marginalized PDFs (diagonal panels) and 2D likelihood contours (lower panels) of the three astrophysical parameters  $\zeta$, log$_{10}(T^{\rm min}_{\rm vir} [K])$ and $R_{\rm mfp}$ (Mpc) for the \textsc{faint sources} model assuming 1000h observations with SKA at redshifts, 10, 9  and 8. The dashed lines show the fiducial parameters ($\zeta$ = 30, log$_{10}(T^{\rm min}_{\rm vir} [K])$ = 4.7 and $R_{\rm mfp}$ = 15). We compare the outcomes of 21CMMC using the power spectrum or the number and morphological spectra of the ionized regions.}
 \label{fig:30_hii_gauss}
\end{figure*}

Figure~\ref{fig:30_hii_gauss} shows the result of the MCMC runs using either the number of ionized structures observed and their morphology or the power spectrum of the 21-cm observations. Additionally, we report in Table~\ref{tab:mcmc_gauss} the median values and  associated 16$^{\rm th}$ and 84$^{\rm th}$ percentile errors. Figure~\ref{fig:30_hii_gauss} shows that the recovered $\zeta$ and  $T^{\rm min}_{\rm vir}$ are shifted towards larger values for both cases. This suggests that the presence of diffuse structures likely suppress the information lying in the small scale structures, such that we recover fewer but brighter sources. Overall, this bias is smaller when using the statistics extracted on 21-cm images. The recovered intervals for $\zeta$ and log$_{10}(T^{\rm min}_{\rm vir})$ are offset by about $+10$ and $+0.1$ compared to the constraints obtained in Section~\ref{sect:resmorph}, and the fiducial values are still within 1.5$\sigma$ of the posterior distributions suggesting that the inferred astrophysics is still consistent with the mock parameters. On the other hand, the intervals recovered using the power spectrum are shifted by more than 3$\sigma$ (+40 and +0.47 for $\zeta$ and log$_{10}(T^{\rm min}_{\rm vir})$, respectively), which could lead to an incorrect astrophysical interpretation. This larger bias can be explained by the fact that the power spectrum is a spherically averaged metric, hence globally sensitive to contamination affecting the contrast of the 21-cm intensities. Typically, adding a GRF to the observation will always bias the power spectrum upward if the field is uncorrelated to the 21-cm field. This is consistent with \citet{Nasirudin2020} who report that including realistic foreground models can bias the parameter constraints up to $5\sigma$, due to the cross-power between the foreground, thermal noise, and EoR signal. On the other hand, a GRF acts like a correlated noise in the tomographic images, with both up and down fluctuations with zero mean, and thus  do not necessarily bias the shape of the PDF of the 21-cm intensities, such that the number and morphological attributes of the ionized regions can still be robustly extracted using the same approach.  

We note that we must remain careful when interpreting the derived uncertainties from the 16$^{\rm th}$ and 84$^{\rm th}$ percentile errors. As mentioned in Section~\ref{sect:likemorph}, the likelihood function for the 21-cm tomographic statistics does not include additional variance terms arising from the expected fluctuations of the morphological attributes due to sample variance or thermal noise. Similarly, including an additional 20\% modelling uncertainty for the power spectrum, as it was done in \citet{greig2015},  might result in larger recover parameter intervals, such that the fiducial values are within 3$\sigma$. Nevertheless, this would not affect the observed biases of the recovered intervals, which itself supports that 21-cm tomographic statistics are likely more robust to the presence of foreground residuals. 

We briefly note that, similar to the results shown in Sections~\ref{sect:resmorph}, \ref{sect:resps} and \ref{sect:sample}, $R_{\rm mfp}$ is still poorly constrained in both cases. This is expected because of the extra level of complexity of this test.

Overall, the interest of 21-cm tomographic statistics to further constrain the EoR is thus well illustrated by this section. These results highlight that they likely provide more accurate, although less precise,  statistics than the power spectrum regarding the presence of potential foreground contamination. This is key information for future observational studies, and we discuss this further in Section~\ref{disc:morphuse}.

\section{Discussion}
\label{sect:disc}
We further discuss here the results presented in Section.~\ref{sect:results}. Section~\ref{sect:indiv} investigates the connection between the morphological pattern spectra of the ionized regions and the astrophysics of the reionization. Section~\ref{disc:morphuse} discusses the potential of 21-cm tomographic statistics for future observational studies. Finally, Section~\ref{disc:limit} considers the limitations of our experiment.

\subsection{The ionized regions morphology probes the average ionizing budget per baryon}
\label{sect:indiv}
\begin{figure*} 
\includegraphics[width = 0.95\textwidth]{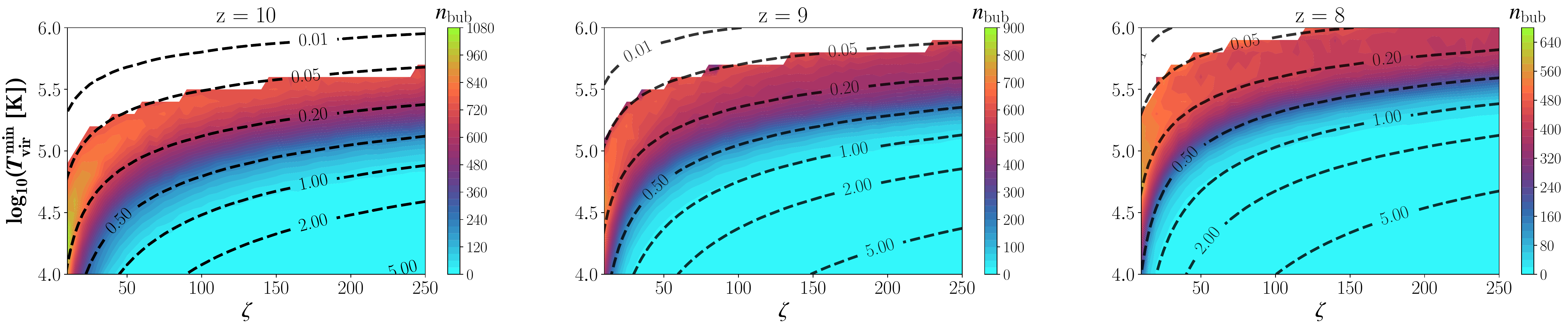}
\includegraphics[width = 0.95\textwidth]{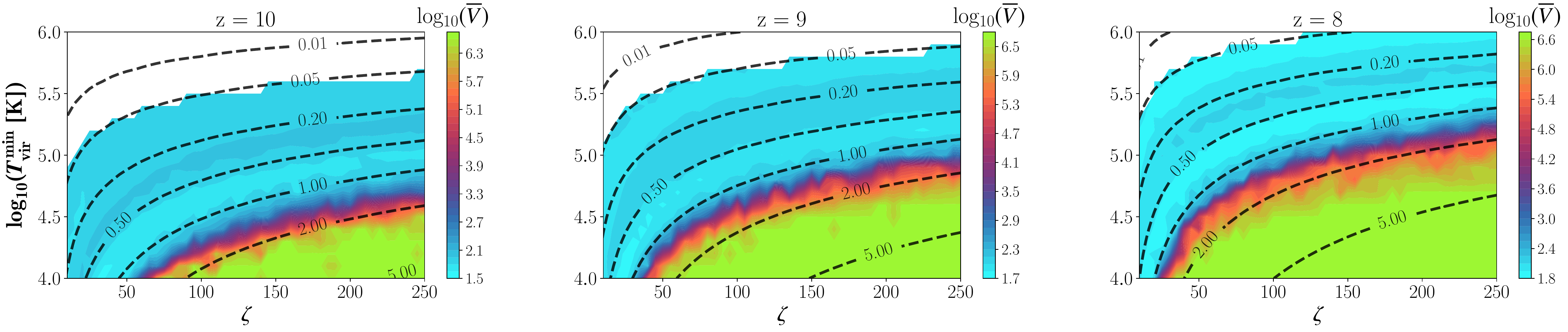}
\includegraphics[width = 0.95\textwidth]{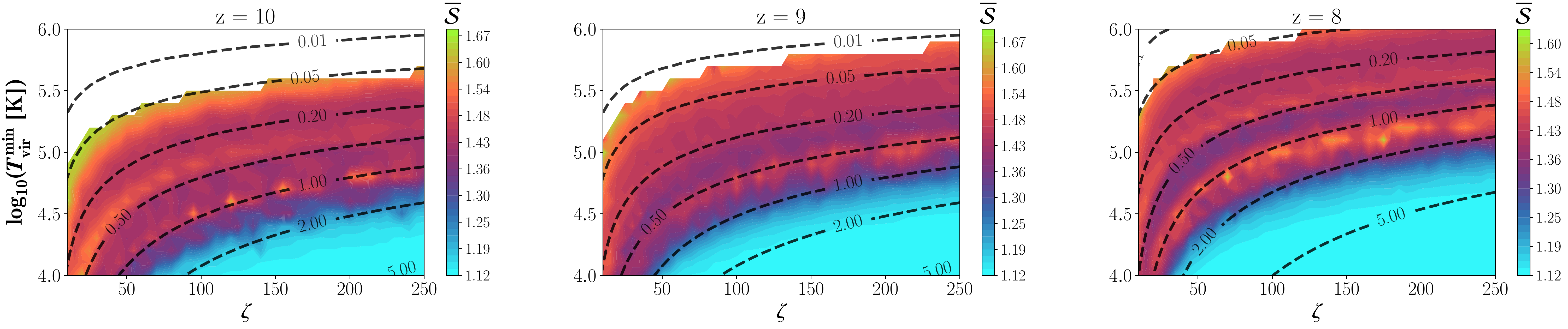}
\includegraphics[width = 0.95\textwidth]{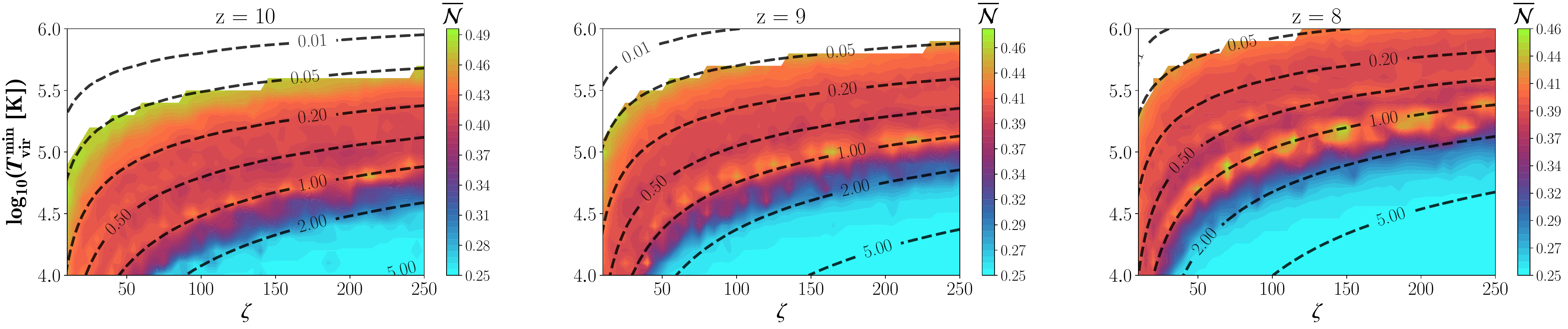}
\includegraphics[width = 0.95\textwidth]{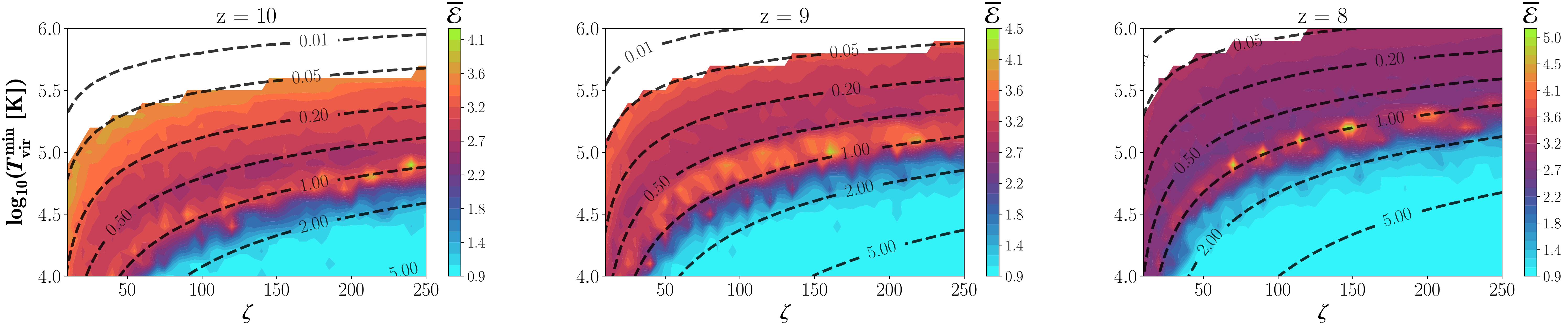}
\includegraphics[width = 0.95\textwidth]{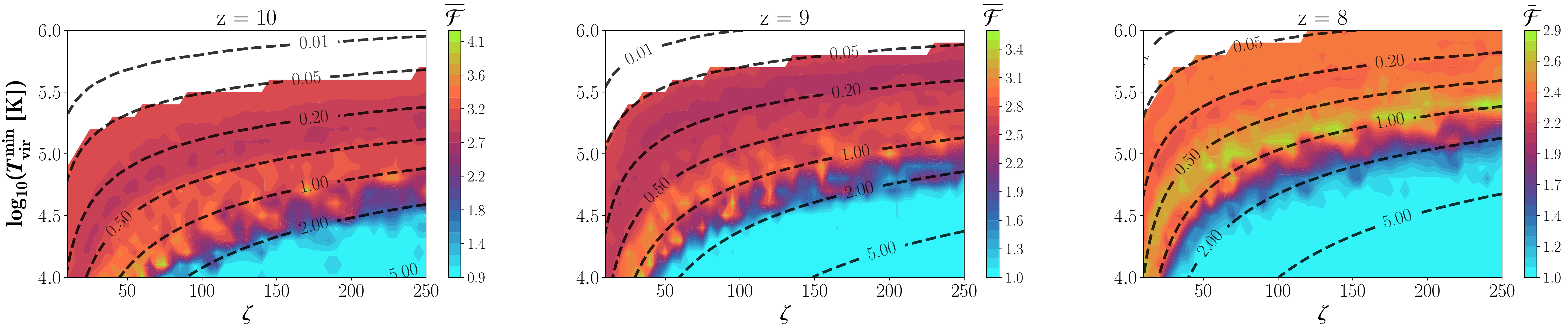}
 \caption{Evolution of the average values of the morphological attributes as a function of the ionizing efficiency and the minimum virial temperature of the star-forming halos, at redshifts 10, 9 and 8. From top to bottom: the number of ionized structures observed; their average volume (logarithmic); sparseness; non-compactness; elongation; and flatness. Each individual plot combines 11871 model regularly sampling the parameter space (see detail in Section~\ref{sect:indiv}). The white regions are models where the ionized fraction is too low to extract the ionized regions (typically $x_{\rm ion} < 0.05$). The black dashed lines corresponds to the isocontours of the average number of ionizing photons emitted per baryon, derived as $\zeta\times f^{\rm P-S}_{\rm coll}({ z}, M_{\rm min})$.}
 \label{fig:colormaps}
\end{figure*}

\begin{figure*} 
\includegraphics[width = 0.8\textwidth]{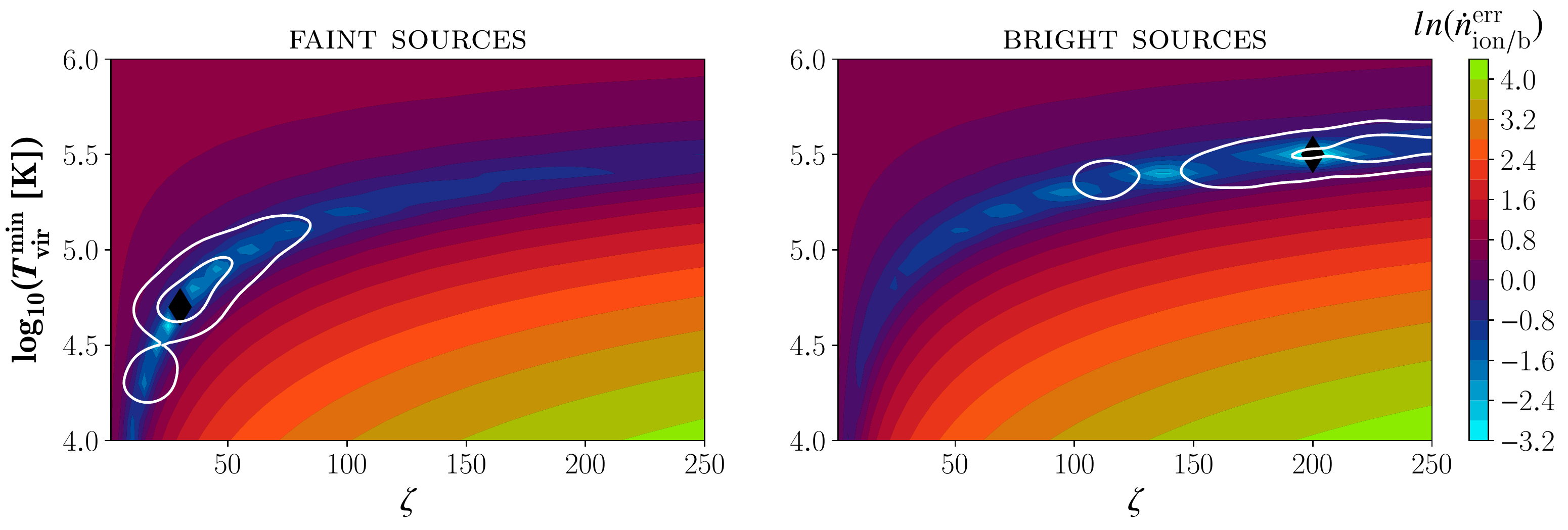}
 \caption{Colormap representing the relative difference ($\dot{n}_{\rm ion/b}^{\rm err}$) of the time evolution of the average number of ionizing photons emitted per baryon, as a function of the model parameters $\zeta$ and log$_{10}(T^{\rm min}_{\rm vir})$ for the \textsc{faint sources} (left panel) and \textsc{bright sources} model (right panel). The white contours show the 2D likelihood contours obtained in Section~\ref{sect:resmorph}.}
 \label{fig:nion}
\end{figure*}

In this section, we investigate which properties, among the number of ionized regions or their morphological attributes, contain most of the constraining power to infer the astrophysical parameters. To do so, we extract these statistics for a large number of models sampled within the parameter space used in this work, and using a grid interval of 5, 0.1, and 2, for $\zeta$, log$_{10}(T^{\rm min}_{\rm vir})$, and $R_{\rm mfp}$, respectively. In total, this corresponds to 11781 models for which we created realistic mock observations at $z$ = 10, 9 and 8, and extracted the ionized regions number count, volume, sparseness, non-compactness, elongation, and flatness. 

Figure~\ref{fig:colormaps} shows the color maps corresponding to these six statistics as a function of $\zeta$ and log$_{10}(T^{\rm min}_{\rm vir})$. To do so, we marginalize the values over $R_{\rm mfp}$ by taking the mean over all 11 $R_{\rm mfp}$ realizations for each pair of $\zeta$ and log$_{10}(T^{\rm min}_{\rm vir})$. Except for the number of ionized regions, we focus on the average of the morphological attributes over all the ionized structures. We note that this is slightly different from the likelihood framework (Eq~\eqref{eq:likemorph}) which compares the distributions in a point by point manner. Additionally, we show on the figure the isocontours (dashed black lines) of the average number of ionizing photons emitted in the IGM per baryon at a given redshift. Typically, the number of ionizing photons emitted in the IGM per baryon, $\dot{n}_{\rm ion/b}$, is a time-dependent quantity, which relates both to the ionizing efficiency of the sources (considered independent from time or halo mass in this work), and the minimum mass of the star-forming halo through the equation:
\begin{equation}
    \dot{n}_{\rm ion/b} = \zeta \times \frac{df_{\rm coll}({z}, M_{\rm min})}{dt}.
\end{equation}
\noindent Hence,  the \textit{average} number of ionizing photons emitted in the IGM per baryon at a given redshift $z$, referred to as $\overline{n}_{{\rm ion/b},\ z}$, can be approximated using the mean collapse fraction in the box (averaged over all voxels). In \textsc{21cmFAST}, the collapse fraction is estimated for different excursion sets corresponding to different scales $R$. Nevertheless, by construction, the average $f_{\rm coll}$ should match the Press-Schechter collapse fraction ($f^{\rm P-S}_{\rm coll}$) defined for a redshift $z$ and a minimum halo mass $M_{\rm min}$. Thus, we derive $\overline{n}_{{\rm ion/b}, z}$ at a given redshift using:
\begin{equation}
    \overline{n}_{{\rm ion/b},\ z} = \zeta \times f^{\rm P-S}_{\rm coll}({ z}, M_{\rm min}).
\end{equation}

\noindent The isocontours in Figure~\ref{fig:colormaps} correspond to the evolution of $\overline{n}_{{\rm ion/b},\ z}$ as a function of $\zeta$ and log$_{10}(T^{\rm min}_{\rm vir})$. The white regions in the different panels are the sets of parameters where the ionized fraction in the box is too low to robustly estimate the number of ionized regions (typically $x_{\rm ion} < 0.05$). These models have large $T^{\rm min}_{\rm vir}$ such that too few ionizing sources are formed and  $\overline{n}_{{\rm ion/b},\ z}$ is less than 0.05. On the other hand, models with $\overline{n}_{{\rm ion/b},\ z} > 1.5$ have few objects, an average volume larger than 10$^6$ Mpc$^3$, and $\overline{\mathcal{E}}$ and $\overline{\mathcal{F}}$ close to 1, suggesting that the universe is dominated by a single percolated ionized region. The models with the most ``extreme" morphological attributes (large $\overline{\mathcal{E}}$, $\overline{\mathcal{F}}$, $\overline{\mathcal{N}}$ or $\overline{\mathcal{S}}$) are located around the isocontour $\overline{n}_{{\rm ion/b},\ z} = 1$. The peculiar morphology of the ionized regions in these models is explained by the fact that this isocontour traces a reionization stage just before percolation. At this point, the ionizing field consist of a few large regions (a few tens in number, with an average volume between $10^2 - 10^3$ Mpc$^3$), strongly porous and eccentric, which are on the brink to fuse into a single, percolated cluster spanning the whole box. 

Overall, Figure~\ref{fig:colormaps} emphasizes that the number of ionized regions and their morphology are strongly correlated with the isocontours of $\overline{n}_{{\rm ion/b}, z}$. Nevertheless, despite a few exceptions, these statistics remain fairly constant along these isocontours, which they are only sufficient to constrain the value of $\overline{n}_{{\rm ion/b}, z}$ that contains the correct astrophysics, but might not be enough to disentangle the correct pair of $\zeta$ and $T^{\rm min}_{\rm vir}$ on this contour. On the other hand, Figure~\ref{fig:colormaps} shows that the shape and location of the isocontours of $\overline{n}_{\rm ion/b}$ evolve with redshift. Hence, combining the information about the ionized regions morphology at different redshifts can indirectly trace the time evolution of the ionizing budget per baryon, and thus provides additional information to recover  $\zeta$ and $T^{\rm min}_{\rm vir}$. We test for this hypothesis by deriving the relative difference in $\dot{n}_{\rm ion/b}$, noted $\dot{n}_{\rm ion/b}^{\rm err}$, for each pair of $(\zeta, T^{\rm min}_{\rm vir})$, relatively to the $\dot{n}_{\rm ion/b}$ of the fiducial models. Hence, $\dot{n}_{\rm ion/b}^{\rm err}$ is defined as:
\begin{equation}
    \dot{n}_{\rm ion/b}^{\rm err}(\zeta, T^{\rm min}_{\rm vir})  = \sqrt{\sum^{ z} \left( \frac{\overline{n}_{{\rm ion/b},\ z}^{\rm \, mock\ model}\ -  \overline{n}_{{\rm ion/b},\ z}(\zeta, T^{\rm min}_{\rm vir})}{\overline{n}_{{\rm ion/b},\ z}^{\rm \, mock\ model}} \right)^2},
\end{equation}

\noindent where $\overline{n}_{{\rm ion/b},\ z}^{\rm \, mock\ model}$ and $\overline{n}_{{\rm ion/b}, z}(\zeta, T^{\rm min}_{\rm vir})$ are the average number of ionizing photons emitted in the IGM per baryon at a given redshift for the fiducial model (\textsc{faint sources} or \textsc{bright sources}), and for a given pair of $\zeta$ and $T^{\rm min}_{\rm vir}$, respectively. Hence, $\dot{n}_{\rm ion/b}^{\rm err}$ traces the difference in the time evolution of the ionizing budget for all the models sampled relative to the fiducial models considered in this work. We present the evolution of $ln(\dot{n}_{\rm ion/b}^{\rm err})$ as a function of  $\zeta$ and $T^{\rm min}_{\rm vir}$ for the \textsc{faint sources} or \textsc{bright sources} cases in Figure~\ref{fig:nion}, and additionally display the 2D likelihood contours obtained for both models in Section~\ref{sect:resmorph} (shown in Figures~\ref{fig:30_hii_morph} and \ref{fig:200_hii_morph}) with black solid lines. We note that the 2D likelihood contours of the inference analysis are consistent with the regions having the lowest $ln(\dot{n}_{\rm ion/b}^{\rm err})$ which characterize the parameter space where the difference in the time evolution of the ionizing budget is minimum. Additionally, the degeneracy between these parameters observed in Section~\ref{sect:resmorph} is directly consistent with the shape of these regions. Therefore, this analysis highlights that the evolution of the ionized regions number count and morphological spectra at different redshifts likely provides the strongest constraints on $\zeta$ and $T^{\rm min}_{\rm vir}$, because it traces $\dot{n}_{\rm ion/b}$. Thus, it supports that these tomographic statistics give robust and independent insights about the physical process of reionization.

\subsection{Combining power spectrum and tomographic analyses}
\label{disc:morphuse}


\citet{koopmans2015} and \citet{mellema2015} pointed out the potential of the synergy between the power spectrum and alternative approaches based on 21-cm tomographic images. Recent studies have further shown that the use of higher-order, topological and morphological statistics could provide valuable additional information to disentangle the astrophysics of the EoR \citep[Section 1 and ][]{greig2019}. Several of them emphasized that these approaches would be especially useful to break the degeneracies between several astrophysical models having a similar power spectrum \citep[e.g.][]{kakiichi2017, giri2018_bsd, kapahtia2019}. In this work, we showed how these 21-cm tomographic statistics could be used in a Bayesian inference framework to disentangle two different reionization scenarios, and that these constraints are theoretically more robust to the presence of residual foregrounds than when using the power spectrum (see Section~\ref{sect:gauss}). Hence, independently combining both approaches can already provide valuable insights into the presence of contamination in the observation if their outcomes are inconsistent.

This first test-case supports the importance of using both approaches to achieve the most robust scientific results and paves the way for further studies investigating the synergy between power spectra and tomographic analyses. In this work, we focused on an intuitive formalism related to the shape or number of ionized regions and emphasized that these statistics trace the time evolution of the ionizing budget per baryon, thus providing already crucial insights on the astrophysics of the EoR. The use of more advanced methods, encoding, for example, a topological analysis of the ionization field \citep{elbers2019} could help to extend this analysis further. Additionally, alternative statistics that only encode the non-gaussianities of the 21-cm signal, such as the bi-spectrum \citep[e.g][]{watkinson2019, hutter2020} or the skewness and kurtosis of the 21-cm PDF \citep{banet2020}, could be more easily combined to the power spectrum within the same MCMC analysis, because they provide independent information. This is less trivial for morphological or topological analysis of the ionizing field, where additional steps would be required to ensure that the information they provide is truly disconnected from the power spectrum.

Hence, further theoretical work is required to investigate the optimal strategy to combine both approaches, to prepare for upcoming observational studies with SKA and HERA.

\subsection{Limitations of using tomography}
\label{disc:limit}
To investigate the use of the 21-cm tomographic statistics to recover the astrophysics of the EoR, we made a few assumptions to either simplify the computation of the 21-cm signal or its analysis. One might thus ask: how dependent are our results from the typical assumptions made? We discuss here the potential caveats arising from these assumptions. 

In this work, we ignored the effects of redshift space distortion (RSD). RSD artificially compresses and expands the high and low-density regions, respectively \citep[Kaiser effect;][]{kaiser1987}. This effect has been studied in \citet{giri2018_bsd}, where the authors showed that it shifts the ionized bubbles size distribution towards smaller values. This is expected from the typical inside-out reionization topology where the ionized regions grow first on the highest density regions, thus are compressed due to the RSD effects. In \citet{giri2018_bsd}, the authors highlight that accounting for these effects has a limited impact on the final results. However, if the data were affected by RSD, they must be included in the simulations to consistently compare the distribution of the number-count and morphological attributes between the observation and the models.

Additionally, our experimental setup considers co-eval cubes for both the mock observations and the models. However, true observations consist of lightcones, such that the line of sight axis can cover a large range of redshifts. The impact of lightcone effects have been discussed several times in the literature \citep[e.g.][]{datta2012, datta2014,greig2018}. \citet{giri2018_bsd} investigated its impact on the ionized bubble size distribution (BSD), and showed that it shifts the ionized regions size towards larger values. However, \citet{datta2014} showed that the impact of lightcone effects can be mitigated by using a limited bandwidth observation, less than $\Delta f = 10$ MHz ($\Delta z$ = 0.5). Thus, similarly to the RSD effects, we do not expect this issue to be a major obstacle for future 21-cm tomographic analyses. 

Another concern is related to the absence of physically motivated foregrounds in our mock observations. Perfect foreground removal is typically assumed in all theoretical papers discussing 21-cm tomographic statistics. This is because, in theory, foregrounds properties are significantly different (e.g smooth, or statistically independent) from the properties of the 21-cm fluctuations. Nevertheless, in practice, perfectly removing all foreground contamination is a very complex task, and artifacts can remain after applying foreground removal methods. In Section~\ref{sect:gauss}, we briefly discussed the possible impact of foreground residual but limited our analysis to a specific case using Gaussian random fields with a given power spectrum shape. This test-case already showed that statistics extracted on 21-cm images are likely more robust to the presence of these artifacts than when using the power spectrum. However, additional studies with more complex physical foregrounds models, are needed to further investigate their impact on power spectra or tomographic analyses. We note however that significant improvements have been made regarding foreground removal algorithms. For example, \citet{chapman2013} used a theoretical experiment based on the LOFAR sensitivity to show that a Generalized Morphological Component Analysis approach \citep[GMCA;][]{bobin2013} would already perform well to recover the 21-cm images from foreground contaminated observations. In fact, \citet{hothi2020} show that the use of Gaussian Regression analysis tools \citep[GPR][]{mertens2018} performs better than GMCA when including the PSF and the noise in the mock observations. Hence the use of already available tools, and the development of futures foreground removal methods, optimal for tomographic applications, will be the key to mitigate the caveats arising from these contaminants. 

Finally, the results presented in this work are strongly dependent on the validity of the $T_{\rm S} >> T_{\rm CMB}$ (post-heating) hypothesis. Indeed, the ionized regions in the 21-cm tomographic images can be robustly extracted, even in the presence of noise or additional smoothing, because they imprint a clear peak in the low-end distribution of the 21-cm intensities. However, this is more complex when the spin temperate is not well heated above the CMB temperature because the presence of low-intensity structures also arise from regions where $T_{\rm S} = T_{\rm CMB}$. Getting rid of this assumption would allow us to probe higher redshifts, constrain epochs where the EoR and the Epoch of Heating overlap \citep[e.g. as in ][]{greig2017}, and more generally, explore more complex reionization models. Nevertheless, it would require to adopt a different approach to probe the evolution of the morphological information lying in the 21-cm fluctuations, for example, by considering the 21-cm field morphological spectra as a function of several threshold sets within the 21-cm intensity range \citep[similar to ][]{kapahtia2019}. Such strategy would be simple to implement using the connected component tree framework of \textsc{DISCCOFAN}, but would require to re-define the likelihood function in \textsc{21CMMC}, which is not a trivial task for this approach. Consequently, dismissing this assumption is likely the more challenging obstacle to overcome to extend this analysis to higher redshifts.

Overall, this work paves the way for further theoretical studies investigating the robustness of 21-cm tomographic statistics using more complex reionization parametrization and observations including both physically motivated foregrounds and additional instrumental artifacts. The ability of 21-cm images to probe the astrophysics, as shown in Section~\ref{sect:resmorph}, and their apparent better robustness to residual contamination (Section~\ref{sect:gauss}), support that they are promising for future observational studies.

\section{Conclusions}
\label{sect:conc}
We have examined the use of statistics extracted from 21-cm tomographic images to constrain the properties of the reionization sources using a Bayesian statistical inference framework. We defined an intuitive formalism based on the number count and morphological pattern spectra of the ionized regions, encoding their eccentricity, porosity and asymmetry. We implemented a likelihood function within \textsc{21CMMC} \citep{greig2015}, a Markov Chain  Monte Carlo analysis tool, to quantify the astrophysical constraints set by these 21-cm tomographic statistics. We used a reionization model with three parameters: the ionizing efficiency of the sources ($\zeta$, assumed constant), the mass of the star-forming halos derived from the minimum virial temperature ($T^{\rm min}_{\rm vir}$), and the mean free path of the ionizing photons in the IGM ($R_{\rm mfp}$). We created mock observations of 128$^3$ cells with 2 Mpc resolution (0.735 arcmin) based on two reionization scenarios dominated by \textsc{faint sources} ($\zeta$ = 30, $T^{\rm min}_{\rm vir} = 5.0\ 10^4$ K, and $R_{\rm mfp}$ = 15 Mpc), and \textsc{bright sources} ($\zeta$ = 200, $T^{\rm min}_{\rm vir} = 3.0\ 10^5$ K, and $R_{\rm mfp}$ = 15 Mpc), similarly as in \citet{greig2017}. We included the noise profile and point spread function of the SKA1-Low, assuming 1000 hours of observation, a maximum baseline of 2 km, and three snapshots at redshift 10, 9,and 8. Our results can be summarized as follow:

\begin{itemize}
    \item The morphological properties of the ionized regions provide valuable information to trace the reionization history (Section~\ref{sect:evolution}) or differences between different reionization scenarios (Section~\ref{sect:model}). Additionally, these morphological differences are more significant in the largest structures, such as the ionized percolated region that typically dominates the late reionization stages. This supports that crucial astrophysical insights can already be extracted from the morphology of a single, large ionized structure.
    \item The ionized regions number-count and morphological spectra, incorporated in a Bayesian inference framework, can disentangle reionization scenarios with different ionizing efficiency of the sources of reionization and minimum mass of the star-forming halos  (Section~\ref{sect:resmorph}). This is because these statistics indirectly probe the time evolution of the average number of ionizing photons emitted in the IGM per baryon, which is fixed for a given pair of $\zeta$ and $T^{\rm min}_{\rm vir}$ (Section~\ref{sect:indiv}). This supports the use of 21-cm tomographic statistics as an alternative approach to the power spectrum to recover the astrophysics of the ionizing sources.
    \item Using the power spectrum of the 21-cm fluctuations provides the tightest constraints for $\zeta$ and $T^{\rm min}_{\rm vir}$, and is slightly less sensitive to different sets of initial conditions than the number and morphology of the ionized regions (Sections~\ref{sect:resps} and \ref{sect:sample}). Nevertheless, statistics extracted from realistic mock 21-cm tomographic observations are more robust to the presence of diffuse foreground residuals in the observations than using the power spectrum (Section~\ref{sect:gauss}). This is shown by the presence of a larger bias in the recovered parameter intervals when using the power spectrum (Figure~\ref{fig:30_hii_gauss}). Comparing the outcomes of power spectra and tomographic analyses is useful to check the consistency of the recovered astrophysical parameters, and can provide valuable information about the presence of residual structures in the observation (Section~\ref{disc:morphuse}).
    \item The mean free path of photons, $R_{\rm mfp}$, can not be accurately inferred using our experimental setup, both when using the power spectrum or the tomographic statistics (Section~\ref{sect:resps}). This is because the mock observations and the models sampled have a limited box size (250 Mpc), such that the large scales, where fluctuations in $R_{\rm mfp}$ have a significant impact, are dominated by statistical variance (Figure~\ref{fig:wrong_ps}). This is supported by the results of Section~\ref{sect:sample}, where we showed that varying the initial conditions has a significant impact on the recovered $R_{\rm mfp}$ value, both for the power spectrum and the tomographic statistics. 
\end{itemize}

This work emphasizes that 21-cm tomographic statistics can be successfully incorporated within a Bayesian statistical inference tool, and provide a robust alternative approach to the power spectrum. Additionally, it highlights the importance of combining both approaches in upcoming observational studies with the SKA to achieve the best scientific results. In further exploratory studies, we plan to assess the robustness of these results using more complex reionization parametrizations, by removing the post-heating assumption ($T_{\rm S} >> T_{\rm CMB}$), and generating mock observations including physically motivated foreground contamination. Overall, this paper supports that the use of 21-cm images and, in general, the morphological properties of the ionized structures, provide key information to the physical process of reionization.

\section*{Acknowledgements}
The authors would like to thank the anonymous referee for constructive comments that help improve the quality of this manuscript, and the Center for Information Technology of the University of Groningen for their support and for providing access to the Peregrine high performance computing cluster. SG thanks Sambit Giri for several discussions throughout this work, Florent Mertens and Garrelt Mellema for helpful suggestions,  Andrei Mesinger and Bradley Greig for discussions about \textsc{21cmFAST} and \textsc{21CMMC}, and Marco Grzegorczyk for an instructive conversation on the implementation of likelihood functions adapted for high-dimensional statistics.  This paper is based on research developed in the DSSC Doctoral Training Programme, co-funded through a Marie Sklodowska-Curie COFUND (DSSC 754315). \\
Additionally to the various tools referenced in the manuscript, the authors acknowledge the use of the python packages  \textsc{numpy} \citep{numpy}, \textsc{matplotlib} \citep{matplotlib}, \textsc{scipy} \citep{scipy}, \textsc{panda} \citep{panda}, and \textsc{scikit-image} \citep{scikit-image} .

\section*{Data availability}
The simulation data underlying this article will be shared on request to the corresponding author.




\bibliographystyle{mnras}
\bibliography{bibliographie.bib} 





\appendix

\section{Comparison of methods to extract the ionized regions from mock tomographic observations}
In Section~\ref{sect:noisepsf}, we briefly mentioned the technique used in this work to extract ionized regions from noisy observations. We provide here more details about the different thresholding methods. Recently, \citet{giri2018_identif} proposed a new approach to optimally identify the ionized regions from realistic observations, using a superpixel thresholding approach, SLIC, introduced in \citet{achanta2012}. While this method shows good performance for their application, the SLIC approach is computationally expensive, which makes it less suitable for Bayesian inference frameworks.

On the contrary, we found the Triangle thresholding approach \citep{zack1977} to perform well on the typical noisy image cubes we deal with in this work. This thresholding method works on the histogram of the values in the image. It constructs a line between the peak and farthest measurement in the tail of the histogram, and the threshold value is derived as the value where the distance between this line and the histogram is maximal. This approach is identical to the maximum deviation method as used by \citet{giri2018_identif}.

\begin{figure*} 
\includegraphics[width=\textwidth]{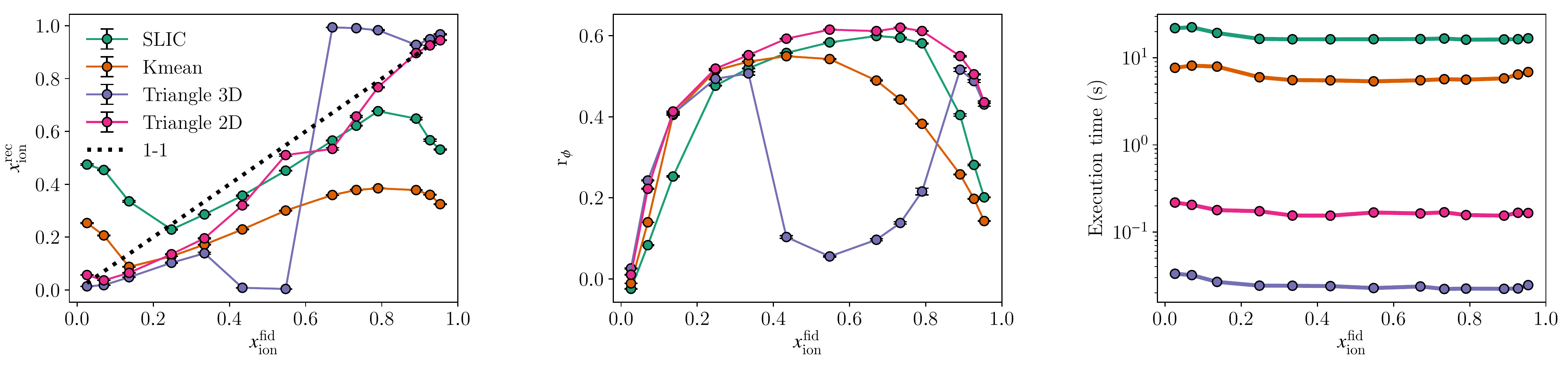}
\caption{Left: the recovered ionizing fraction in the realistic observations versus its fiducial value. Middle: The correlation coefficient between the recovered binary fields using the different methods, and the ionization field extracted from the fiducial simulation. Right: The computational time of each method.}
\label{figapp:threshcomp}
\end{figure*}

In Figure~\ref{figapp:threshcomp}, we compare the performance of four approaches, the SLIC method, the k-means clustering used for example in \citet{kakiichi2017}, and the Triangle thresholding approach. We note that we applied the latter method in two different ways, labeled as Triangle3D and Triangle2D. Triangle3D refers to the case where we apply the triangular thresholding  directly on the three-dimensional image. Triangle2D refers to a different approach where we first apply the triangular thresholding to all individual 2D slices and collect the corresponding list of threshold values. We then take the median value of this list and use the latter to segment the entire three-dimensional 21-cm map in a consistent manner. This approach is different from stacking the set of 2D individually thresholded slices and does not bias the recovered spatial extent of the ionized regions in the resulting 3D segmented volume. To compare the performance of these methods, we used several co-eval cubes of 128$^3$ cells with $\approx 2$ Mpc resolution, with different neutral fractions in the box. We added a noise cube corresponding to 1000 hours observations with SKA1-Low at $z$ = 10, and smoothed the final cube assuming a resolution corresponding to maximum baseline of 2 km (the rms of the noise after smoothing is 2.5 mK). The left panel compares the recovered volumetric ionizing fraction $x^{\rm rec}_{\rm ion}$ with respect to the fiducial ionizing fraction $x^{\rm fid}_{\rm ion}$ in the box. The middle panel shows the evolution of the correlation coefficient \citep[$r_\phi$][]{cramer1946} between the recovered binary fields using the different methods, and the binary field from the fiducial simulations extracted from the ionization fraction box taking $x^{\rm fid}_{\rm ion} > 0.5$ as a threshold to determine if a voxel is ionized or not. The right panel shows the execution time to extract the ionized regions as a function of the fiducial $x_{\rm ion}$.

Overall, the correlation coefficient shows that the Triangle2D approach performs better than the other methods, while its execution time is lower than 0.1 second. Interestingly, the triangle thresholding approach that finds the best on the 2D slices and taking the median works better than applying it directly on the 3D data. SLIC performs slightly better than Triangle2D, but is two order of magnitude slower. Additionally, the triangle method has a similar maximum deviation method inWe note that the performance of SLIC is not as good as in \citet{giri2018_identif}, where the authors report e.g a $r_\phi$ of 0.6 at $x^{\rm fid}_{ion}$ = 0.9, while we derived 
$r_\phi$ = 0.45 for the same ionizing fraction. However, the authors have used different box size (250$^3$ cells with 1.38 cMpC resolution), hence results are not directly comparable. Therefore, the triangular thresholding approach is more adapted for our framework, but we urge caution about generalizing these results to other applications.

\label{app:thresh}

\section{Power spectrum of the mock observations}
\label{app:ps}
\begin{figure*} 
\includegraphics[width=\textwidth]{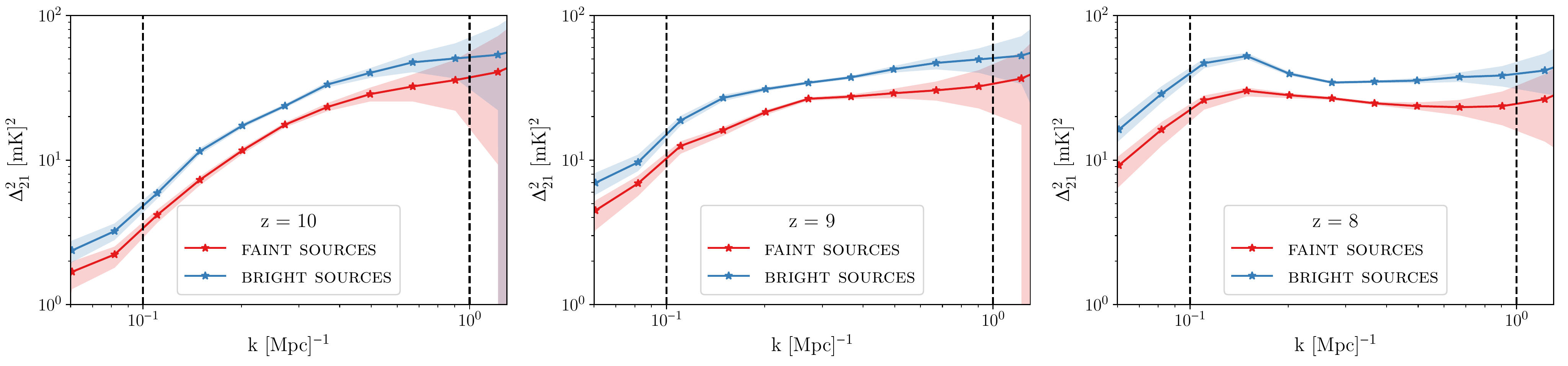}
\caption{The 21-cm power spectrum for the \textsc{faint sources} (red solid line) and \textsc{bright sources}  (blue solid line) mock observations at the three co-eval redshifts used in this work. Red and blue shaded area correspond to the noise uncertainty (thermal + sample variance) corresponding to 1000 hours of observation with SKA. }
\label{figapp:ps}
\end{figure*}

Figure~\ref{figapp:ps} shows the 21-cm power spectrum of the \textsc{faint sources}  ($\zeta$ = 30, log$_{10}(T^{\rm min}_{\rm vir} [K])$ = 4.7 and $R_{\rm mfp}$ = 15 Mpc) and \textsc{bright sources} ($\zeta$ = 200, log$_{10}(T^{\rm min}_{\rm vir} [K])$ = 5.48 and $R_{\rm mfp}$ = 15 Mpc)  mock observations \citep{mesinger2016} at the three co-eval redshifts used in this work. The total uncertainty on each power spectrum is shown by the shaded areas around the power spectra, and is based on 1000 hours of observations with SKA1-Low, derived using the method described in Section~\ref{sect:likeps}. As seen in this figure, the theoretical uncertainty is particularly small for most of the $k$ bins between 0.1 and 0.4 Mpc$^{-1}$. This suggests that differences between several reionization models are particularly significant at the large scales (low $k$), and that this information can be used to disentangle different reionization scenarios. However, as discussed in Section~\ref{sect:gauss}, the presence of contaminants, such as residual diffuse structures in the observation, can have a strong impact on the observed power spectrum, biasing it upward if the structures are uncorrelated to the 21-cm field.

\section{Likelihood for N-dimensional distributions}
\label{app:like}

In this section, we provide a general framework to define an approximate likelihood estimator suitable to compare two distributions of $n$ vectors in a N-dimensional space, with N typically larger than 3. This formalism accounts for the fact that $n$ might fluctuate for different observations, and its fluctuations relate to the parameters of the models to be inferred. We define two distributions A and B, with $n_{\rm A}$ and $n_{\rm B}$ observed structures, respectively. The first factor of the likelihood function is a Poissonian factor that compares $n_{\rm A}$ to $n_{\rm B}$, assuming that $n_{\rm A}$ is the number of structures observed, and $n_{\rm B}$ is the expectation value derived from the model:
\begin{equation}
    \frac{\exp{(-n_{\rm B})}\ \times\  n_{\rm  B}^{n_{\rm A}} }{ n_{\rm A}!} 
\end{equation}

The second factor must provide a similarity metric to compare the distribution of N-dimensional vectors from A and B. The strategy consists in finding, for each vector $\vec{v}_{\rm A}$ in A, the most similar vector $\vec{v}_{\rm B}$ in the distribution B.  This can be achieved using the Mahalanobis distance metric, \citep{mahalanobis1936}, which is particularly suited to compare points from N-dimensional distributions \citet{aggarwal2001}. We then sum over all the minimum Mahalanobis distances for the vectors in the distributions A, which can be expressed as $\sum^{n_{\rm A}}_i \argminA_{j \in\ n_{\rm  B}} ({\rm d_{\rm maha}^{\rm N}} (\vec{v}_{{\rm A}, i},\vec{v}_{{\rm B},j}))$. Intuitively, this can be understood as looking how closely the distribution B represents the observations in A. Since the Poisson factor accounts for fluctuations in the number of structures in A and B, we need to ensure that this second factor is independent to the number of structures observed in the data and model. To picture this normalization problem, we consider the factor $\rm d^{\rm N}_{maha}$ as a volume in a N-dimensional space. The total volume occupied by the observations in B should be divided by $n_{\rm B}$, such that the volume per observation in the distribution is proportional to $1 / n_{\rm B}$. Thus, if $n_{\rm B}$ gets larger than $n_{\rm A}$, the sum of $d^{\rm N}_{maha}$, which encodes the minimum volume between the observations in A and B, will reach lower values because of $1 / n_{\rm B}$ dependence. To compensate for this effect, we multiply the sum of $\rm d^{\rm N}_{maha}$ by $n_{\rm B}$ such that the final factor is independent from $n_{\rm B}$. Similarly, we add a factor $1 / n_{\rm A}$ to ensures that the sum of the distances is also normalized with respect with the number of objects in the distribution A. Hence, this second factor, referred to as the \textit{distance} factor, is expressed as:


\begin{equation}  
\exp{\left( - \frac{n_{\rm  B}}{n_{\rm A}}\sum^{n_{\rm A}}_i \argminA_{j \in\ n_{\rm  B}} {\rm d_{\rm maha}} (\vec{v}_{{\rm A}, i},\vec{v}_{{\rm B},j}))^{\rm N}\right) }.
\end{equation}

\noindent We additionally include a regularization factor $\lambda$ to adjust the relative importance of each term, such that, larger and smaller $\lambda$ values respectively increase and decrease the weight on the factor that compares the N-dimensional vectors. The final expression of our maximum likelihood estimator function is defined as:
\begin{multline}  
\mathcal{L}(\text{morphology}) = \frac{ \exp{(-n_{\rm B})}\ \times\  n_{\rm  B}^{n_{\rm A}} }{n_{\rm A}!} \\ \times \exp{\left( - \lambda \times \frac{n_{\rm  B}}{n_{\rm A}}\sum^{n_{\rm A}}_i \argminA_{j \in\ n_{\rm  B}} {\rm d_{\rm maha}} (\vec{v}_{{\rm A}, i},\vec{v}_{{\rm B},j}))^{\rm N}\right) }.
\end{multline}

We note that this equation is not  strictly speaking an exact likelihood function. It suffers a few caveats, as it is not normalized, and does not encode potential systematics related to the variance of the measurement of the N-dimensional vectors in A and B. The presence of the regularization factor $\lambda$ makes it  closer to a regularized function. However, it provides a suitable framework for comparing the distributions of $n$   N-dimensional vectors, where $n$ can significantly fluctuate, and additionally includes the information lying in the distribution outliers, which are typically not accounted for in classical distribution metrics (e.g. mean or standard deviation). This strategy can be less effective when N is large and the vectors of observations extracted from A or B are subject to noise, because distributions with outliers will be penalized by the factor $d^{\rm N}_{maha}$.  However, we also tested different cases using $d^{\rm K}_{maha}$, with K < N, to reduce this caveat. We found that the best results were achieved when K = N.



\section{Sample variance tests}
\label{app:tablemcmc}
In this appendix, we detail the individual results of the ten \textsc{21CMMC} tests with different initial conditions done in Section~\ref{sect:sample}. Table~\ref{tab:mcmc_sample} shows the individual median, 16$^{\rm th}$ and 84$^{\rm th}$ percentile errors for the three astrophysical parameters, $\zeta$,  log$_{10}(T^{\rm min}_{\rm vir})$, and $R_{\rm mfp}$ considering the \textsc{faint sources} model, and two cases using either the power spectrum or the number and morphological properties of the ionized regions.

\begin{table*}
	\centering

	\label{tab:mcmc_sample}
	\begin{tabular}{ccccc} 
		\hline
		Initial conditions set & Case & $\zeta$ & log$_{10}(T^{\rm min}_{\rm vir} [K])$ & $R_{\rm mfp}$ (Mpc)\\
		\hline		\hline
		\multirow{2}{*}{1} &Ionized regions count + morphology & $35.02^{+18.36}_{-7.39}$ & $4.78^{+0.19}_{-0.12}$ & $12.54^{+5.26}_{-2.84}$\\
        & Power spectrum & $31.29^{+1.24}_{-1.17}$ & $4.74^{+0.03}_{-0.03}$ & $22.11^{+1.76}_{-2.31}$\\\\
        
		\multirow{2}{*}{2} &Ionized regions count + morphology & $31.80^{+34.43}_{-11.31}$ & $4.71^{+0.39}_{-0.25}$ & $13.55^{+5.31}_{-4.70}$\\
        & Power spectrum & $31.11^{+1.29}_{-1.21}$ & $4.72^{+0.03}_{-0.03}$ & $18.24^{+3.42}_{-1.80}$\\\\
        
		\multirow{2}{*}{3} &Ionized regions count + morphology & $47.38^{+17.17}_{-6.90}$ & $4.96^{+0.12}_{-0.07}$ & $15.88^{+4.22}_{-3.80}$\\
        & Power spectrum & $30.09^{+1.48}_{-1.27}$ & $4.71^{+0.04}_{-0.03}$ & $14.83^{+2.56}_{-1.09}$\\\\
        
		\multirow{2}{*}{4} &Ionized regions count + morphology & $39.32^{+29.14}_{-8.40}$ & $4.90^{+0.20}_{-0.12}$ & $17.14^{+2.98}_{-4.78}$\\
        & Power spectrum & $30.90^{+1.28}_{-1.17}$ & $4.70^{+0.03}_{-0.03}$ & $21.65^{+1.99}_{-2.32}$\\\\
        
		\multirow{2}{*}{5} &Ionized regions count + morphology & $29.90^{+57.80}_{-2.16}$ & $4.70^{+0.49}_{-0.04}$ & $14.07^{+4.03}_{-3.14}$\\
        & Power spectrum & $29.65^{+1.13}_{-1.06}$ & $4.66^{+0.03}_{-0.02}$ & $20.12^{+2.55}_{-2.35}$\\\\
        
		\multirow{2}{*}{6} &Ionized regions count + morphology & $28.28^{+2.91}_{-3.56}$ & $4.72^{+0.06}_{-0.07}$ & $15.66^{+5.21}_{-2.90}$\\
        & Power spectrum & $30.05^{+1.23}_{-1.10}$ & $4.70^{+0.03}_{-0.03}$ & $20.34^{+2.24}_{-2.12}$\\\\
        
		\multirow{2}{*}{7} &Ionized regions count + morphology & $37.26^{+8.88}_{-8.09}$ & $4.82^{+0.09}_{-0.12}$ & $16.25^{+3.29}_{-3.50}$\\
        & Power spectrum & $30.64^{+1.26}_{-1.14}$ & $4.73^{+0.03}_{-0.03}$ & $18.61^{+2.25}_{-1.70}$\\\\
        
		\multirow{2}{*}{8} &Ionized regions count + morphology & $31.31^{+21.36}_{-9.82}$ & $4.73^{+0.23}_{-0.27}$ & $15.50^{+3.51}_{-2.31}$\\
        & Power spectrum & $30.48^{+1.32}_{-1.25}$ & $4.70^{+0.03}_{-0.03}$ & $18.01^{+4.41}_{-3.18}$\\\\
        
		\multirow{2}{*}{9} &Ionized regions count + morphology & $43.06^{+11.80}_{-11.50}$ & $4.88^{+0.10}_{-0.18}$ & $15.87^{+4.53}_{-3.58}$\\
        & Power spectrum & $30.37^{+1.23}_{-1.19}$ & $4.71^{+0.03}_{-0.03}$ & $18.57^{+3.03}_{-2.42}$\\\\
        
		\multirow{2}{*}{10} &Ionized regions count + morphology & $35.91^{+48.42}_{-18.11}$ & $4.83^{+0.38}_{-0.44}$ & $16.51^{+4.14}_{-4.31}$\\
        & Power spectrum & $29.87^{+1.16}_{-1.08}$ & $4.69^{+0.03}_{-0.03}$ & $18.58^{+3.18}_{-2.00}$\\\\
        \hline
		\multirow{2}{*}{Average} &Ionized regions count + morphology & $36.92^{+26.73}_{-15.04}$ & $4.82^{+0.24}_{-0.26}$ & $15.53^{+4.16}_{-4.29}$\\
        & Power spectrum & $30.24^{+3.34}_{-3.34}$ & $4.70^{+0.05}_{-0.04}$ & $19.29^{+3.35}_{-3.22}$\\\\
	\end{tabular}
		\caption{The median values, and the associated 16$^{\rm th}$ and 84$^{\rm th}$ percentile errors of the three recovered astrophysical parameters intervals, using either the power spectrum or the number and morphology of the ionized regions. We ran ten different MCMC setups using different sets of initial conditions, and the mock \textsc{faint sources} observation,  assuming 1000h of observations with SKA1-Low for three different epochs at $z$ = 10, 9 and 8. Additionally, the last row shows the constraints and intervals derived from the average 1D PDFs of all ten realizations. }
\end{table*}

\bsp	
\label{lastpage}
\end{document}